# FORCE-DETECTED NUCLEAR MAGNETIC RESONANCE

M. POGGIO AND B. E. HERZOG

DEPARTMENT OF PHYSICS

UNIVERSITY OF BASEL

KLINGELBERGSTRASSE 82

4056 BASEL, SWITZERLAND

## 1 INTRODUCTION

The drive to improve the sensitivity of nuclear magnetic resonance (NMR) to smaller and smaller sample volumes has led to the development of a variety of techniques distinct from conventional inductive detection. In this chapter, we focus on the technique of force-detected NMR as one of the most successful in yielding sensitivity improvements. We review the rationale for the technique, its basic principles, and give a brief history of its most important results. We then cover in greater detail its application in the first demonstration of three-dimensional (3D) nuclear magnetic resonance imaging (MRI) with nanometer-scale resolution. Next, we present recent developments and likely paths for improvement. Finally, the technique and its potential are discussed in the context of competing and complementary technologies.

## 2 MOTIVATION

In 1981 Binnig, Gerber, and Weibel introduced the scanning tunneling microscope (STM) [Binnig1982], which – for the first time – provided real-space images of individual atoms on a surface. The closely related invention of the atomic force microscope (AFM) by Binnig [BinnigPat1986] and its subsequent realization by Binnig, Quate, and Gerber [BinnigPRL1986], both in 1986, eventually expanded atomic-scale imaging to a wide variety of surfaces beyond the conducting materials made possible by STM. The key component of an AFM is its force sensor, which is a transducer used to convert force into displacement, i.e. a spring, coupled with a sensitive optical or electrical displacement detector. Although early AFM transducers were simply pieces of gold or aluminum foil [BinnigPat1986, Rugar1990], specially designed and mass-produced Si cantilevers soon became the industry standard and led to improved resolution and force sensitivity [Akamine1990]. These micro-processed devices are now cheap, readily available, and designed – depending on the target application – to have integrated tips and a variety of other features including coatings or electrical contacts.

It is in the midst of these developments in the 1980s and early 1990s that modern force-detected NMR was born. As scanning probe microscopy (SPM) expanded its applications to magnetic force microscopy (MFM), Sidles proposed a force microscopy based on magnetic resonance as a method to improve the resolution of MRI to molecular length-scales [Sidles1991,SidlesPRL1992]. Soon after the proposal in 1991, Rugar realized magnetic resonance force microscopy (MRFM) by using an AFM cantilever to first detect electron spin resonance (ESR) in 1992 [Rugar1992] and then NMR in 1994 [Rugar1994].

Prompted by the rapid progress and astounding success of SPM in achieving atomic-scale imaging of surfaces, a number of researchers set about adapting these advances to the problem of magnetic resonance imaging (MRI). This work was motivated by the visionary goal of imaging molecules atom-by-atom, so as to directly map the 3D atomic structure of macromolecules [SidlesRSI1992]. The realization of such a "molecular structure microscope" would have a dramatic impact on modern structural biology, and would be an important tool for many future nanoscale technologies. While the ultimate goal of atomic-scale MRI still remains unachieved today, MRFM has undergone a remarkable development into one of the most sensitive magnetic resonance methods available to researchers today. Among the important experimental achievements are the detection of a single electronic spin [Rugar2004] and the extension of the spatial resolution of nuclear MRI from several micrometers to below ten nanometers [Degen2009].

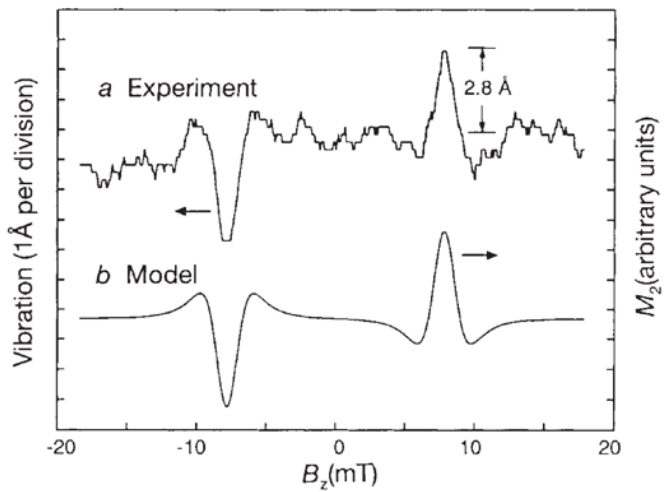

Figure 1: ESR signal from the first demonstration of MRFM by Rugar et al. in 1992 [Rugar1992].

## 3 PRINCIPLE

In conventional nuclear magnetic resonance detection, the sample is placed in a strong static magnetic field in order to produce a Zeeman splitting between spin states. The sample is then exposed to an RF magnetic field of a precisely defined frequency. If this frequency matches the Zeeman splitting, then the system absorbs energy from the RF radiation resulting in transitions between the spin states. The resulting oscillations of this ensemble of magnetic moments produce a time-varying magnetic signal that can be detected with a pickup coil. The electric current induced in the coil is then amplified and converted into a signal that is proportional to the number of moments (or spins) in the sample. In MRI, this signal can be reconstructed into a 3D image of the sample using spatially varying magnetic fields and Fourier transform techniques. The magnetic fields produced by nuclear moments are, however, extremely small: more than $10^{12}$ nuclear spins are typically needed to generate a detectable signal.

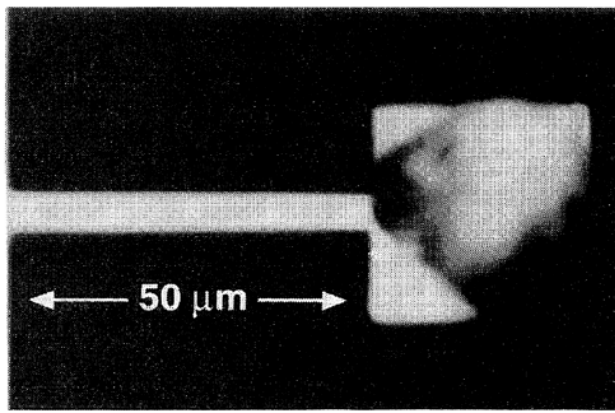

Figure 2: Optical micrograph of a 90 nm thick silicon nitride cantilever with a sample of ammonium nitrate attached used in the first demonstration of nuclear MRFM by Rugar et al. in 1994 [Rugar1994].

MRFM relies on the mechanical measurement of the weak magnetic force between a microscopic magnet and the magnetic moments in a sample. These moments are due to either the atomic nuclei with nonzero nuclear spin or electron spins present in a sample. For a single magnetic moment $\boldsymbol{\mu}$ in a magnetic field $\boldsymbol{B}$, this force can be expressed as:

$$\boldsymbol{F} = \nabla(\boldsymbol{\mu} \cdot \boldsymbol{B}). \tag{1}$$

Using a compliant cantilever, one can measure the component of $\boldsymbol{F}$ along the cantilever's deflection direction $\hat{x}$:

$$F_x = \frac{\partial}{\partial x}(\boldsymbol{\mu} \cdot \boldsymbol{B}) = \mu \frac{\partial B_z}{\partial x} = \mu\, G, \tag{2}$$

where $\boldsymbol{\mu}$ points along $\hat{z}$ and $G = \frac{\partial B_z}{\partial x}$ is a magnetic field gradient. First, either the sample containing nuclear or electronic moments or the nano-magnet must be fixed to the cantilever. The sample and magnet must be in close proximity, sometimes up to a few tens of nanometers from each other. A nearby radio-frequency (RF) source produces magnetic field pulses similar to those used in conventional MRI, causing the moments to periodically flip. This periodic inversion generates an oscillating magnetic force acting on the cantilever. In order to resonantly excite the cantilever, the magnetic moments must be inverted at the cantilever's mechanical resonance frequency. The cantilever's mechanical oscillations are then measured by an optical interferometer or beam deflection detector. The electronic signal produced by the optical detector is proportional to the cantilever oscillation amplitude, which depends on the number of moments in the imaging volume. Spatial resolution results from the fact that the nanomagnet produces a magnetic field which is a strong function of position. The magnetic resonance condition and therefore the region in which the spins periodically flip is confined to a thin, approximately hemispherical "resonant slice" that extends outward from the nano-magnet, as shown in Fig. 3. By scanning the sample in 3D through this resonant region, a spatial map of the magnetic moment density can be made. Different types of magnetic moments (e.g., $^{1}$H, $^{13}$C, $^{19}$F, or even electrons) can be distinguished due to their different magnetic resonance frequencies, giving an additional chemical contrast.

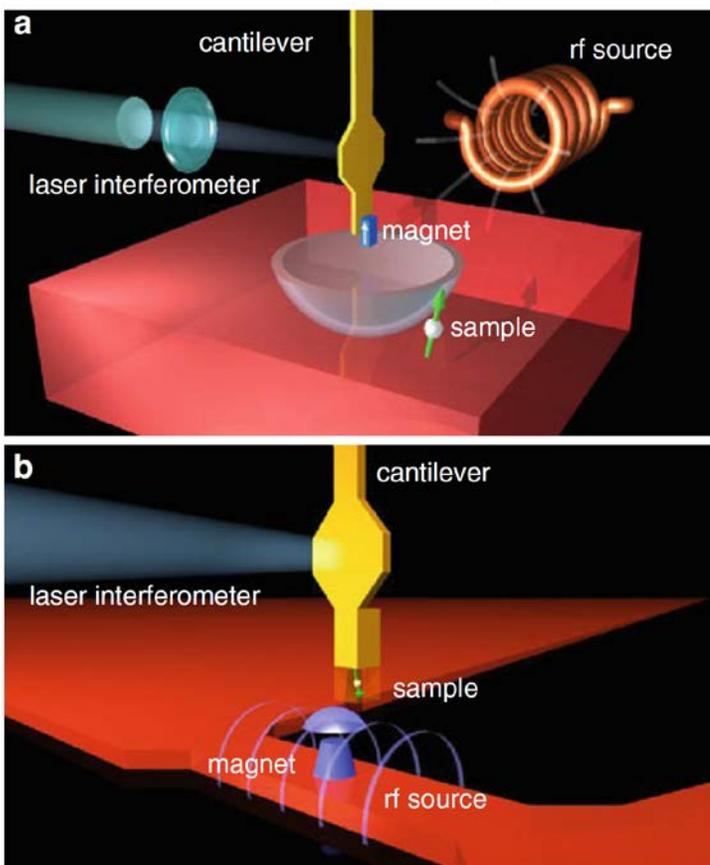

Figure 3: Schematics of an MRFM apparatus. (a) Corresponds to the "magnet-on-cantilever" arrangement, such as used in the single-electron MRFM experiment of 2004 [9]. (b) Corresponds to the "sample-on-cantilever" arrangement, like the one used for the nanoscale virus imaging experiment in 2009 [Degen2009].

## 4 FORCE VS. INDUCTIVE DETECTION

In order to understand why force-detected NMR is well-suited to small sample volumes, we go back to the analysis of Sidles and Rugar [Sidles1993]. In their 1993 letter, they compare inductive and mechanical methods for detecting magnetic resonance. They consider both detection setups as oscillators coupled to a spatially localized magnetic moment. In the first case, the oscillator is an electrical LC circuit – the pick-up coil – inductively coupled to the magnetic moment. In the second case, the oscillator is a mechanical spring – the cantilever – holding the magnetic moment, which is coupled to the field gradient of a nearby magnet. The two cases turn out to be mathematically identical and can be characterized by three parameters: an angular resonance frequency $\omega_0$, a quality factor $Q$, and a "magnetic spring constant" $k_m$ with units of J/T², which is defined in a way that both the electrical and mechanical oscillators are treated on the same footing. Intuitively this quantity can be understood, for the coil, as the energy required to produce an oscillating field within its volume. For the cantilever, it is the energy required to produce the same oscillating field within the sample by moving it in the magnet's field gradient. The authors show that the signal-to-noise ratio of the two magnetic resonance detection schemes is proportional to:

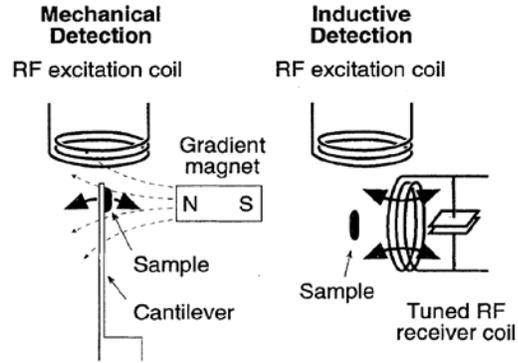

Figure 4: Mechanical vs. inductive detection of magnetic resonance as discussed in Sidles and Rugar [Sidles1993].

$$SNR \propto \sqrt{\frac{\omega_0 Q}{k_m}}. \quad (3)$$

For conventional inductive detection with a cylindrical coil, $k_m$ is proportional to the volume of the coil; for force detection $k_m$ depends on the magnetic field gradient and the size and aspect ratio of the cantilever: $k_m \propto G \frac{w\,t^3}{l^3}$, where $w$, $t$, and $l$ are the width, thickness, and length of the cantilever, respectively. The minute dimensions and extreme aspect ratios of cantilevers as well as the strong micro- and nanometer-scale magnets routinely realized by modern fabrication techniques ensure that $k_m$ is much smaller for modern force-detected techniques than for inductively-detected techniques. An MRFM apparatus using a cantilever with a spring constant of 50 μN/m and a magnetic tip with field gradient of 5 × 10⁶ T/m has $k_m = 2 \times 10^{-18}$ J/T²; a small coil of 4 turns with a diameter of 1.8 mm and a length of 3 mm has has $k_m = 1.2 \times 10^{-2}$ J/T² [Moores2016]. Intuitively one can understand this huge disparity by considering that producing an oscillating field within the whole volume of an inductive pick-up coil can easily require more energy than moving a tiny sample on a compliant cantilever through a magnetic field gradient.

Note that although $\omega_0$ is typically above 100 MHz for inductive detection and in the few kHz range for many force-detected schemes, the difference in $k_m$ of practically achievable coils and cantilevers more than compensates. In addition, mechanical devices usually have a quality factor $Q$ that surpasses that of inductive circuits by orders of magnitude, resulting in a much lower baseline noise. For example, state-of-the art cantilever force transducers achieve $Q$ between 10⁴ and 10⁷, enabling the detection of forces of aN/(Hz)^(1/2) – less than a billionth of the force needed to break a single chemical bond. In addition, scanning probe microscopy offers the stability to position and image samples with nanometer precision. The combination of these features allows mechanically detected MRI to image at resolutions that are far below 1 μm and – in principle – to aspire to atomic resolution

For sensitive transducers, experiments show that $Q$ is limited by surface-related losses, which leads to a linear decrease with increasing surface-to-volume ratio, i.e. $Q \propto t$ [TaoNano2015]. Furthermore, given that for a cantilever $\omega_0 \propto \frac{t}{l^2}$, if we fix the transducer's aspect ratio and shrink each of its dimensions (i.e. multiply each dimension by a factor $\epsilon < 1$ ), $\sqrt{\frac{\omega_0 Q}{k_m}}$ will increase with the square root ( $\sqrt{\frac{\omega_0 Q}{k_m}} \propto \epsilon^{-1/2}$). The result is in an increase in signal-to-noise proportional to the square root of the shrinkage ($SNR \propto \epsilon^{-1/2}$). Given the advent of bottom-up synthesis [Poggio2013], ever smaller mechanical devices are becoming possible making force-detected NMR a potentially ideal technique for pushing towards ever greater sensitivity and smaller detection volumes.

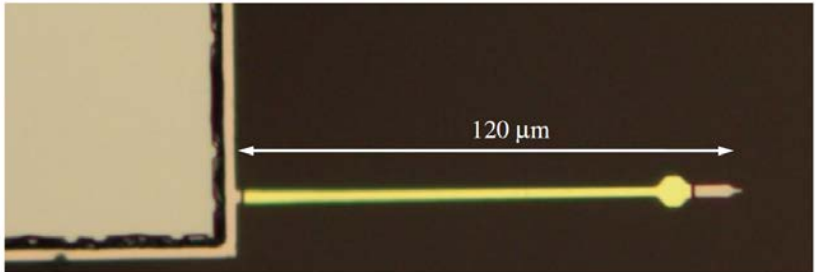

**Figure 5: Optical micrograph of an ultraensitive 100-nm-thick Si cantilever with a spring constant under 100 µN/m. This type of cantilever is used in the most sensitive MRFM experiments to date [Degen2009].**

Although similar scaling arguments can be made for the miniaturization of inductive coils – even resulting in a signal-to-noise increase proportional to the square of the shrinkage ($SNR \propto \epsilon^{-2}$) – the potential gains are more modest given that the technique has less room for improvement. The most sensitive pick-up coils are already close to their practical limits with lengths and diameters around 100 µm (similar to the 20 µm diameter of the wire itself). These nearly optimal coils still have signal-to-noise ratios much smaller [Ciobanu2002] than recent force-detected methods [Degen2009]. In addition, signal-to-noise gains, which can be made by increasing $\omega_0$, are limited by practically achievable laboratory magnetic fields, which have plateaued in recent years around a $^1$H Larmor frequency of 1 GHz [Zalesskiy2014].

## 5 EARLY FORCE-DETECTED MAGNETIC RESONANCE

Force-detection techniques in NMR experiments date back to Evans in 1956 [Evans1956], and were also used in paramagnetic resonance measurements by Alzetta et al. in the 1960s [Alzetta1967]. Sidles' 1991 proposal that magnetic resonance detection and imaging with atomic resolution could be achieved using microfabricated cantilevers and nanoscale ferromagnets [Sidles1991] came, as discussed previously, after the invention of the STM and AFM and in the midst of the rapid progress that followed. Rugar realized the first micrometer-scale experiment using cantilevers [Rugar1992], demonstrating mechanically detected ESR in a 30 ng sample of diphenylpicrylhydrazyl (DPPH), as shown in Fig. 1. The original apparatus operated in vacuum and at room temperature with the DPPH sample attached to the cantilever. A mm-sized coil produced an RF magnetic field tuned to the electron spin resonance of the DPPH at 220 MHz with a magnitude of 1 mT. The electron spin magnetization in the DPPH was modulated by varying the strength of an 8 mT polarizing magnetic field in time. A nearby NdFeB magnet produced a magnetic field gradient of 60 T/m, which, as a consequence of the sample's oscillating magnetization, resulted in a time-varying force between the sample and the magnet. This force modulation was converted into mechanical vibration by the compliant cantilever. Displacement oscillations were detected by a fiber-optic interferometer achieving a thermally limited force sensitivity of 3 fN/(Hz)$^{1/2}$.

Following this initial demonstration of cantilever-based MRFM, the technique has undergone a series of developments towards higher sensitives that, as of today, is 7 orders of magnitude better that of the 1992 experiment [Poggio2010]. Nevertheless, the basic idea of detecting magnetic resonance using a compliant cantilever and a strong magnetic field gradient persists. We now briefly review the important steps that led to these advances while also touching on the application of the technique to imaging and magnetic resonance spectroscopy. Several review articles and other book chapters have appeared that discuss some of these earlier steps more broadly and in richer detail [Nestle2001, Suter2004, Berman2006, Hammel2007, KuehnJCP2006, Barbic2009].

Two years after Rugar's initial demonstration of mechanically-detected ESR, he employed a similar scheme for NMR of a micrometer-scale ammonium nitrate sample shown in Fig. 2 [Rugar1994]. In 1996, Zhang et al. used the technique to detect ferromagnetic resonance (FMR) in a micrometer-scale yttrium iron garnet (YIG) film [Zhang1996]. The first major step towards higher sensitivity was made by incorporating the MRFM instrument into a cryogenic apparatus in order to reduce the thermal force noise of the cantilever. A first experiment carried out in 1996 at a temperature of 14 K achieved a force sensitivity of 80 aN/(Hz)$^{-1/2}$ [Wago1996], a roughly 50-fold improvement compared to 1992, mostly due to the higher cantilever mechanical quality factor and the reduced thermal noise achieved at low temperatures. In 1998, researchers introduced the "magnet-on-cantilever" scheme [WagoAPL1998], where the roles of gradient magnet and sample were interchanged. Using this approach, field gradients of up to $2.5 \times 10^5$ T/m were obtained by using a magnetized sphere of 3.4 μm diameter [Bruland1998]. These gradients were more than 3 orders of magnitude larger than those achieved in the first MRFM experiment. At the same time, a series of spin-detection protocols were also invented. These protocols include the detection of spin signals in the form of a shift in the cantilever resonance frequency (rather than changes in its oscillation amplitude) [StipePRL2001A], and a scheme that relies on detecting a force-gradient, rather than the force itself [Garner2004]. In 2003, researchers approached the level of sensitivity necessary to measure statistical fluctuations in small ensembles of electron spins, a phenomenon that had previously only been observed with long averaging times [Mamin2003]. Further refinements finally led to the demonstration of single electron spin detection in 2004 by the IBM group [Rugar2004].

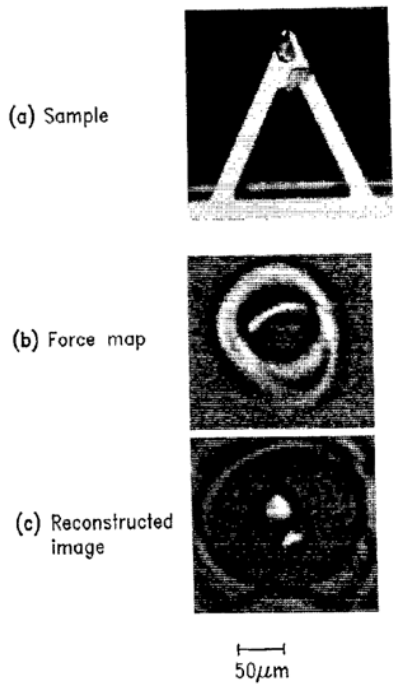

**Figure 6:** (a) Optical micrograph showing two DPPH particles attached to a silicon nitride cantilever. (b) Magnetic resonance force map of the sample. (c) Reconstructed spin density obtained by deconvolving the data in (b) [Zuger1993].

In addition to steady advances in sensitivity, researchers also pushed the capabilities of MRFM for imaging. The first one-dimensional MRFM image was made using ESR detection in 1993 and soon after was extended to two and three dimensions [Zuger1993, Zuger1994, Zuger1996]. These experiments reached about 1 μm axial and 5 μm lateral spatial resolution, which is roughly on par with the best conventional ESR microscopy experiments today [Blank2003]. In 2003, sub-micrometer resolution (170 nm in one dimension) was demonstrated with NMR on optically pumped GaAs [Thurber2003]. In parallel, researchers started applying the technique for the 3D imaging of biological samples, like the liposome, at micrometer resolutions [Tsuji2004]. Shortly thereafter, an 80 nm voxel size was achieved in an ESR experiment that introduced an iterative 3D image reconstruction technique [Wago1997]. The one-dimensional imaging resolution of the single electron spin experiment in 2004, finally, was about 25 nm [8].

The prospect of applying the MRFM technique to nanoscale spectroscopic analysis has also led to efforts towards combination with pulsed NMR and ESR techniques. MRFM is ill suited to high-resolution spectroscopy as broadening of resonance lines by the strong field gradient of the magnetic tip completely dominates any intrinsic spectral features. Nevertheless, a number of advances have been made. In 1997, MRFM experiments carried out on phosphorus-doped silicon were able to observe the hyperfine splitting in the EPR spectrum [Wago1997]. Roughly at the same time, a series of basic pulsed magnetic resonance schemes were demonstrated to work well with MRFM, including spin nutation, spin echo, and $T_1$ and $T_{1\rho}$ measurements [Schaff197, WagoPRB1998]. In 2002, researchers applied nutation spectroscopy to quadrupolar nuclei in order to extract local information on the quadrupole interaction [Verhagen2002]. This work was followed by a line of experiments that demonstrated various forms of NMR spectroscopy and contrast, invoking dipolar couplings [Degen2005], cross polarization [Lin2006, Eberhardt2007], chemical shifts [Eberhardt2008], and multidimensional spectroscopy [Eberhardt2008]. Some interesting variants of MRFM that operate in homogeneous magnetic fields were also explored. These techniques include measurement of torque rather than force [Alzetta1967, Ascoli1996] and the so-called "Boomerang" experiment [Leskowitz1998, Madsen2004].

More recently, experiments in which magnetic field gradients can be quickly switched on and off, have again raised the possibility of doing high resolution spectroscopy by MRFM. In 2012, Nichol et al. realized nuclear MRFM of $^1$H in nanometer-scale polystyrene sample using a nanowire (NW) transducer and a nanometer-scale metallic constriction in order to produce both the RF field and a switchable magnetic field gradient [Nichol2012]. In 2015, Tao et al. demonstrated the use of a commercial hard disk write head for the production of large switchable gradients in an MRFM apparatus [TaoArxiv2015]. These innovations will be discussed in further detail in the section on possible future paths for improvement.

Finally, while not within the scope of this review, it is worth mentioning that MRFM has also been successfully applied to a number of ferromagnetic resonances studies, in particular for probing the resonance structure of micron-sized magnetic disks [Wigen2006, deLoubens2007].

## 6 SINGLE-ELECTRON MRFM

The first decade of MRFM development concluded with the measurement of a single electron spin by the IBM group in 2004. The apparatus combined many of the advances made in the previous years and stands out as one of the first single-spin measurements in solid-state physics. The exceptional measurement sensitivity required for single-spin detection was enabled by several factors, including the operation of the apparatus at cryogenic temperatures and high vacuum, the ion-beam-milling of magnetic tips in order to produce large gradients, and the fabrication of mass-loaded attonewton-sensitive cantilevers [Chui2003], as shown in Fig. 5. The thermal noise in higher order vibrational modes of mass-loaded cantilevers is suppressed compared with the noise in the higher order modes of conventional, "flat" cantilevers. Since high-frequency vibrational noise in combination with a magnetic field gradient can disturb the electron spin, the mass-loaded levers proved to be a crucial advance for single electron MRFM. In addition, the IBM group developed a sensitive interferometer employing only a few nanowatts of optical power for the detection of cantilever displacement [Mamin2001]. This low incident laser power is crucial for achieving low cantilever temperatures and thus minimizing the effects of thermal force noise. A low-background measurement protocol called OSCAR based on the NMR technique of adiabatic rapid passage was also employed [StipePRL2001B]. Finally, the experiment required the construction of an extremely stable measurement system capable of continuously measuring for several days in an experiment whose single-shot signal-to-noise ratio was just 0.06 [Rugar2004].

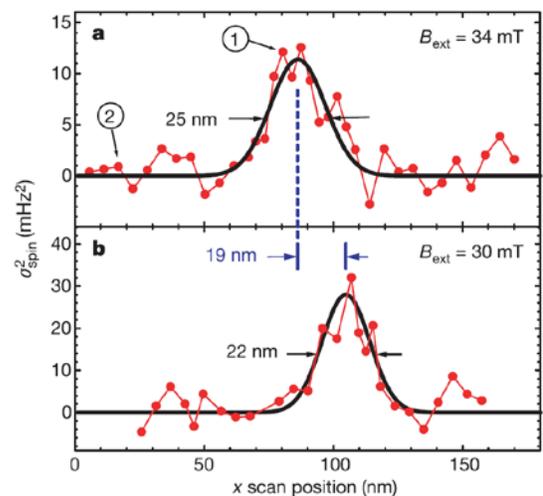

Figure 7: Spin signal as the sample was scanned laterally in the x-direction for two values of external field. The smooth curves are Gaussian fits that serve as guides to the eye. The 19-nm shift in peak position reflects the movement of the resonant slice induced by the 4-mT change in external field [Rugar2004].

The path to this experimental milestone led through a variety of interesting phenomena. In 2003, for example, researchers reported on the detection and manipulation of small ensembles of electron spins—ensembles so small that their statistical fluctuations dominate the polarization signal [Mamin2003]. The approach developed for measuring statistical polarizations provided a potential solution to one of the fundamental challenges of performing magnetic resonance experiments on small numbers of spins. In 2005, Budakian took these concepts one step further by actively modifying the statistics of the naturally occurring fluctuations of spin polarization [Budakian2005]. In one experiment, the researchers polarized the spin system by selectively capturing the transient spin order. In a second experiment, they demonstrated that spin fluctuations can be rectified through the application of real-time feedback to the entire spin ensemble.

## 7 TOWARDS NANO-MRI WITH NUCLEAR SPINS

While the impressive sensitivity gains made by MRFM in mechanically detected ESR demonstrated the technique's promise, the ultimate goal of mapping atomic structure of samples in 3D requires the detection of single nuclear spins. Nuclear MRI has had a revolutionary impact on the field of non-invasive medical screening and is finding an increased number of applications in materials science and biology. The realization of MRI with nanometer or sub-nanometer resolution may have a similar impact, for example, in the field of structural biology. Using such a technique, it may be possible to image complex biological structures, even down to the scale of individual molecules, revealing features not elucidated by other methods.

As a consequence, in the last decade, researchers have focused their efforts on nuclear spin detection by MRFM. The detection of a single nuclear spin, however, is far more challenging than that of single electron spin. This is because the magnetic moment of a nucleus is much smaller: a $^1$H nucleus (proton), for example, possesses a magnetic moment that is only ~1/650 of an electron spin moment. Other important nuclei, like $^{13}$C or a variety of isotopes present in semiconductors, have even weaker magnetic moments. In order to observe single nuclear spins, it is necessary to improve the state-of-the-art sensitivity by another two to three orders of magnitude. While not out of the question, this is a daunting task that requires significant advances to all aspects of the MRFM technique.

### 7.1 IMPROVEMENTS TO MICRO-FABRICATED COMPONENTS

Improvements in the sensitivity and resolution of mechanically detected MRI hinge on a simple signal-to-noise ratio, which is given by the ratio of the magnetic force power exerted on the cantilever over the force noise power of the cantilever device. For small volumes of spins, statistical spin polarizations are measured, therefore force powers (or variances) are of interest rather than force amplitudes:

$$SNR_{MRFM} = N \frac{(\mu_N G)^2}{S_F \Delta f}. \tag{4}$$

Here, $N$ is the number of spins in the detection volume, $\mu_N$ is the magnetic moment of the nucleus of interest, $G$ is the magnetic field gradient at the position of the sample, $S_F$ is the force noise spectral density set by the fluctuations of the cantilever sensor, and $\Delta f$ is the bandwidth of the measurement, determined by the nuclear spin relaxation rate. This expression gives the single-shot signal-to-noise ratio of a thermally limited MRFM apparatus. The larger this signal-to-noise ratio is, the better the spin sensitivity will be.

From the four parameters appearing in (4), only two can be controlled and possibly improved. On the one hand, the magnetic field gradient $G$ can be enhanced by using higher quality magnetic tips and by bringing the sample closer to these tips. On the other hand, the force noise spectral density $S_F$ can be reduced by going to lower temperatures and by making intrinsically more sensitive mechanical transducers. Continued improvements to MRFM sensitivity rely on advances made to both of these critical parameters.

## 7.2 MRI WITH RESOLUTION BETTER THAN 100 NM

In 2007, the IBM group introduced a micro-machined array of Si cones as a template and deposited a multilayer Fe/CoFe/Ru film to fabricate nanoscale magnetic tips [Mamin2007]. The micro-machined tips produced magnetic field gradients in excess of $10^6$ T/m owing to their sharpness (the tip radius is less than 50 nm). Previously, maximum gradients of $2 \times 10^5$ T/m had been achieved by ion-beam-milling SmCo particles down to 150 nm in size. Mamin et al. used a "sample-on-cantilever" geometry with a patterned 80 nm thick $CaF_2$ film as their sample. The $CaF_2$ films were thermally evaporated onto the end of the cantilever and then patterned using a focused ion beam, creating features with dimensions between 50 and 300 nm. The cantilevers used in these measurements were custom-made single-crystal Si cantilevers with a 60 μN/m spring constant and a force sensitivity of around 1 aN/Hz$^{1/2}$ at 1 K [Chui2003].

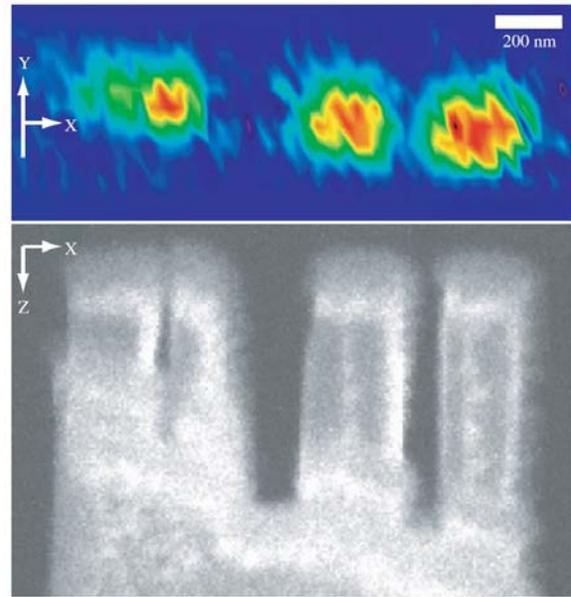

Figure 8: 2D MRFM image of $^{19}$F nuclear spins in a patterned $CaF_2$ sample, and (b) corresponding SEM micrograph (side view) of the cantilever end with the 80 nm thin $CaF_2$ film at the top of the image [Mamin2007].

Fig. 8 shows the result of such an imaging experiment, measuring the $^{19}$F nuclei in the $CaF_2$ sample. The resultant image reproduced the morphology of the $CaF_2$ sample, which consisted of several islands of material, roughly 200 nm wide and 80 nm thick, at a lateral resolution of 90 nm. At a temperature of 600 mK and after 10 min of averaging, the achieved detection sensitivity (SNR of 1) corresponded to the magnetization of about 1200 $^{19}$F nuclear moments.

## 7.3 NANOSCALE MRI OF VIRUS PARTICLES

In the two following years, the group made further improvements to their measurement sensitivity through the development of a magnetic tip integrated onto an efficient "microwire" RF source [Poggio2007], illustrated in Fig. 9. This change in the apparatus solved a simple but significant problem: the typical solenoid coils used to generate the strong RF pulses for spin manipulation dissipate large amounts of power, which even for very small microcoils with a diameter of 300 μm amounts to over 0.2 W. This large amount of heat is far greater than the cooling power of available dilution refrigerators. As a result, nuclear spin MRFM experiments had to be performed at elevated temperatures (4 K or higher), thereby degrading the SNR. In some cases, the effects can be mitigated through pulse protocols with reduced duty cycles [Garner2004, Mamin2007], but it is desirable to avoid the heating issue altogether.

Micro-striplines, on the other hand, can be made with sub-micrometer dimensions using e-beam lithography techniques. Due to the small size, the stripline confines the RF field to a much smaller volume and causes minimal heat dissipation. Using e-beam lithography and lift-off, the IBM group fabricated a Cu "microwire" device that was 0.2 µm thick, 2.6 µm long, and 1.0 µm wide. A stencil-based process was then used to deposit a 200 nm diameter FeCo tip on top of the wire to provide a static magnetic field gradient. Since the sample could be placed within 100 nm of the microwire and magnetic tip, RF magnetic fields of over 4 mT could be generated at 115 MHz with less than 350 µW of dissipated power. As a result, the cantilever temperature during continuous RF irradiation could be stabilized below 1 K, limited by other experimental factors and not the RF device. Simultaneously, the cylindrical geometry of the magnetic tip optimized the lateral field gradient as compared to the micro-machined thin-film Si tips, resulting in values exceeding $4 \times 10^6$ T/m. As an added benefit, the alignment of the apparatus was simplified as the magnetic tip and the RF source were integrated on a single chip. The cantilever carrying the sample simply needed to be positioned directly above the microwire device. Previous experiments had required an involved three-part alignment of magnetic-tipped cantilever, sample, and RF source.

Following the introduction of the integrated microwire and tip device, the IBM researchers were able to improve imaging resolutions to well below 10 nm [Degen2009]. These experiments, which used single tobacco mosaic virus (TMV) particles as the sample, both show the feasibility for MRI imaging with nanometer resolution, and the applicability of MRFM to biologically relevant samples.

Fig. 10 is a representation of the MRFM apparatus used in these experiments. The virus particles were transferred to the cantilever end by dipping the tip of the cantilever into a droplet of aqueous solution containing suspended TMV. As a result, some TMV were attached to the gold layer previously deposited on the cantilever end. The density of TMV on the gold layer was low enough that individual particles could be isolated. Then the cantilever was mounted into the low-temperature, ultra-high-vacuum measurement system and aligned over the microwire.

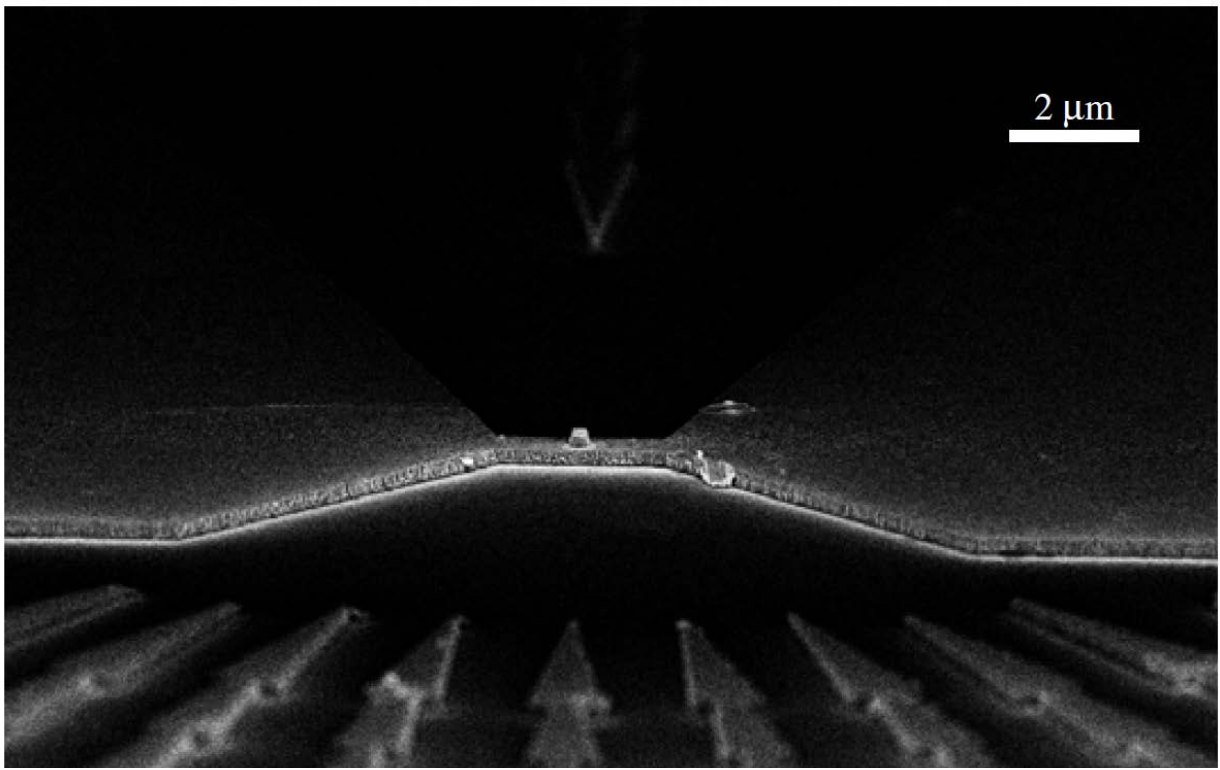

**Figure 9: A SEM of a Cu "microwire" RF source with integrated FeCo tip for MRFM [Poggio2007].**

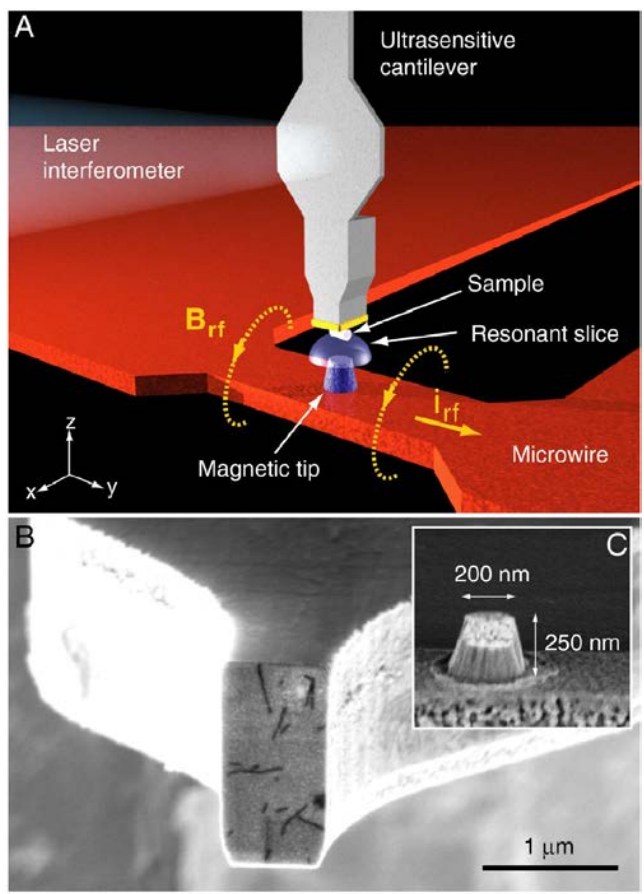

Figure 10: MRFM apparatus. (A) TMV particles, attached to the end of an ultrasensitive silicon cantilever, are positioned close to a magnetic tip. A RF current passing through a copper microwire generates an alternating magnetic field that induces magnetic resonance in the $^1$H spins of the virus particles. The resonant slice represents those points in space where the field from the magnetic tip (plus an external field) matches the condition for magnetic resonance. 3D scanning of the tip with respect to the cantilever, followed by image reconstruction is used to generate a 3D image of the spin density in the virus sample. (B) Scanning electron micrograph of the end of the cantilever. Individual TMV particles are visible as long, dark rods on the sample platform. (C) Detail of the magnetic tip [Degen2009].

After applying a static magnetic field of about 3 T, resonant RF pulses were applied to the microwire source in order to flip the $^1$H nuclear spins at the cantilever's mechanical resonance. Finally, the end of the cantilever was mechanically scanned in three dimensions over the magnetic tip. Given the extended geometry of the region in which the resonant condition is met, i.e. the "resonant slice", a spatial scan does not directly produce a map of the $^1$H distribution in the sample. Instead, each data point in the scan contains force signal from $^1$H spins at a variety of different positions. In order to reconstruct the three-dimensional spin density (the MRI image), the force map must be deconvolved by the point spread function (PSF) defined by the resonant slice. Fortunately, this point spread function can be accurately determined using a magneto-static model based on the physical geometry of the magnetic tip and the tip magnetization. Deconvolution of the force map into the three-dimensional $^1$H spin density can be done in several different ways; for the results presented in [Degen2009] the authors applied the iterative Landweber deconvolution procedure suggested in an earlier MRFM experiment [Chao2004, Dobigeon2009]. This iterative algorithm starts with an initial estimate for the spin density of the object and then improves the estimate successively by minimizing the difference between the measured and predicted spin signal maps. The iterations proceed until the residual error becomes comparable with the measurement noise.

The result of a representative experiment is shown in Fig. 11. Here, clear features of individual TMV particles, which are cylindrical, roughly 300 nm long, and 18 nm in diameter, are visible. As is often the case, both whole virus particles and particle fragments are observed. Given that the raw MRFM data are spatially under-sampled and have only modest SNR, the quality of the reconstruction is remarkable. The observation of significant improvement in image SNR after reconstruction is expected because most spins contribute force signal to more than one position in the scan, and the cumulative effect benefits the SNR of the reconstruction. The resolution appears to be in the 4- to 10-nm range, depending on the direction, with the x-direction having the highest resolution. This resolution anisotropy is expected because of the directional dependence of the PSF, which reflects the fact that the cantilever responds only to the x-component of magnetic force.

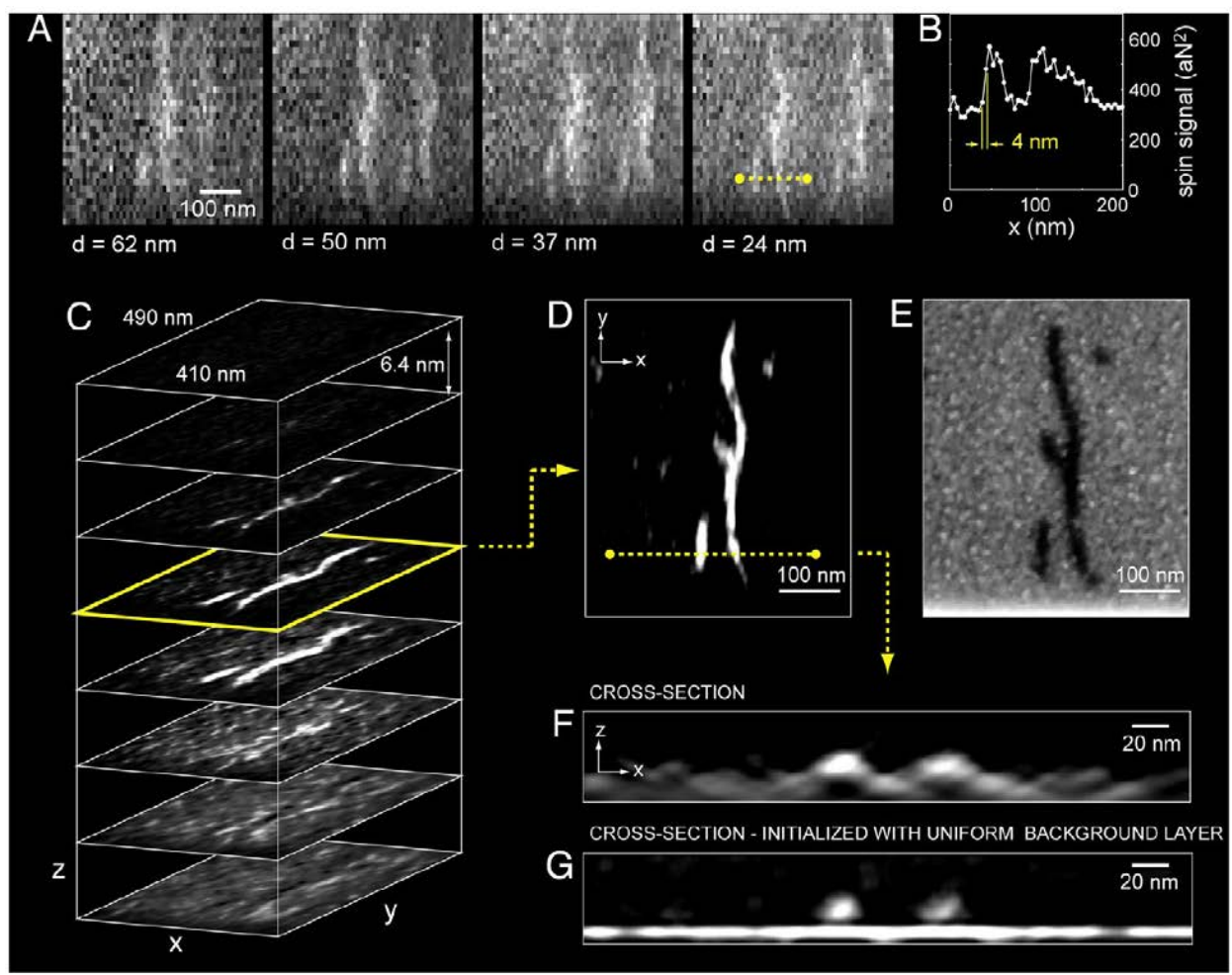

**Figure 11:** Raw data and resulting 3D reconstruction of the $^1$H density distribution. (A) Raw scan data presented as xy-scans of the spin signal at 4 different tip-sample spacings. Pixel spacing is 8.3 nm × 16.6 nm in x × y, respectively. Each data point represents the spin signal variance obtained during a 1-min integration. (B) A more finely sampled line scan showing 4-nm lateral resolution. The scanned region is indicated by the dashed line in A. (C) Reconstructed 3D $^1$H spin density. Black represents very low or zero density of hydrogen, whereas white is high hydrogen density. The image is the result of the Landweber reconstruction, followed by a 5-nm smoothing filter. (D) Horizontal slice of C, showing several TMV fragments. (E) Scanning electron micrograph of the same region. (F) Cross-section showing 2 TMV particles on top of a hydrogen-rich background layer adsorbed on the Au surface. (G) Reconstruction is improved if this background layer is included as a priori information by assuming a thin, uniform plane of 1H density as the starting point of the reconstruction [Degen2009].

The fidelity of the MRFM reconstruction is confirmed by comparing the results to the SEM image of the same sample region in Fig. 11E. Excellent agreement is found even down to small details. Note that the origin of contrast in MRFM image and the SEM image is very different: the MRFM reconstruction is elementally specific and shows the 3D distribution of hydrogen in the sample; contrast in the SEM image is mainly due to the virus blocking secondary electrons emitted from the underlying gold-coated cantilever surface. In fact, the SEM image had to be taken after the MRFM image as exposure to the electron beam destroys the virus particles. The imaging resolution, while not fine enough to discern any internal structure of the virus particles, constitutes a 1000-fold improvement over conventional MRI, and a corresponding improvement of volume sensitivity by about 100 million.

### 7.4 IMAGING ORGANIC NANOLAYERS

In addition to "seeing" individual viruses, the researchers also detected an underlying proton-rich layer. This signal originated from a naturally occurring, sub-nanometer thick layer of adsorbed water and/or hydrocarbon molecules.

The hydrogen-containing adsorbates picked up on a freshly cleaned gold surface turn out to be enough to produce a distinguishable and characteristic signal. From analysis of the signal magnitude and magnetic field dependence, the scientists were able to determine that the adsorbates form a uniform layer on the gold surface with a thickness of roughly 0.5-1.0 nm [Mamin2009].

Using a similar approach, Mamin et al. made a 3D image of a multi-walled nanotube roughly 10 nm in diameter, depicted in Fig. 12. The nanotube, attached to the end of a 100 nm thick Si cantilever, protruded a few hundred nanometers from the end of the cantilever. As had been previously observed with gold layers, the nanotube was covered by a naturally occurring proton-containing adsorption layer. Though the magnitude of the signal was roughly ten times less than that of the two-dimensional layer—reflecting its relatively small volume—it was accompanied by a very low-background noise level that made it possible to produce a clear image of the morphology of the nanotube. Using the same iterative deconvolution scheme developed to reconstruct the image of the TMV particles, the researchers produced an image of a cylindrical object, 10 nm in diameter at the distal end. No evidence was found for the hollow structure that might be expected in the image of such a layer. Given the small inner diameter (less than 10 nm), however, it was not clear whether hydrogen-containing material was in fact incorporated inside the nanotube, or if the resolution of the image was simply not sufficient to resolve the feature.

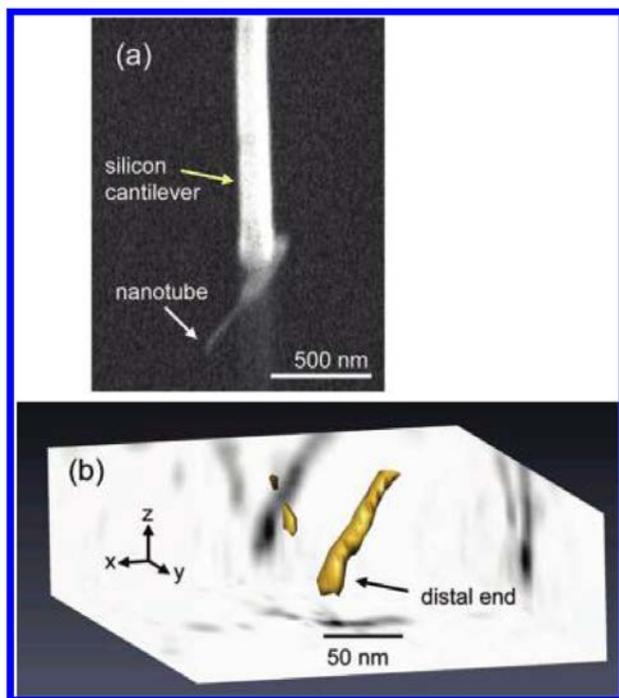

Figure 12: (a) SEM of multi-walled carbon nanotube (NT) attached to a silicon cantilever (side view). The thin NT is supported by a thicker NT that was affixed to the cantilever and then thickened further via electron-beam deposited contamination. (b) 3D image reconstructed from a 3D MRFM data set. The rendered object represents a surface of constant $^1$H density [Mamin2009].

### 8 PATHS TOWARD CONTINUED IMPROVEMENT

Since its invention and early experimental demonstration in the 1990s [Sidles1991, Rugar192, Rugar1994], MRFM has progressed in its magnetic sensitivity from the equivalent of $10^9$ to presently about 100 $^1$H magnetic moments. In order to eventually detect single nuclear spins and to image molecules at atomic resolution, the signal-to-noise ratio of the measurement must still improve by two orders of magnitude. It is not clear if these advances can be achieved by incremental progress to the key components of the instrument, i.e. cantilever force transducers and nanoscale magnetic tips, or whether major shifts in instrumentation and methodology will be necessary.

Since 2009, no further improvements in resolution have been demonstrated beyond the level achieved in the TMV experiment. Nevertheless, extremely promising steps have been taken in the form of both incremental improvements to components and demonstrations of major changes to the measurement technique. From (4), we know that there are essentially two experimental parameters that can be improved: 1) the magnetic field gradient $G$ and 2) the cantilever force noise spectral density $S_F$. From an experimental point of view, in the first case, the task translates to either improving the gradient source, i.e. the magnetic tip, or reducing the surface-induced force noise so that the sample can be brought closer to the magnetic tip. In the second case, the challenge is to optimize the cantilever transducer by reducing intrinsic sources of mechanical dissipation. Finally, we add a third possibility for improvement: development of new measurement protocols, e.g. Fourier encoding or hyper-polarization, which can also lead to gains in SNR for a fixed integration time.

In the following we review experiments occurring since 2009, which all broadly fall into one of the aforementioned categories for improvement of MRFM sensitivity.

## 8.1 MAGNETIC FIELD GRADIENTS

The magnetic force on the cantilever can be enhanced by increasing the magnetic field gradient $G$. This is achieved by making higher quality magnetic tips with sharp features and high-moment materials, and by simultaneously bringing the sample closer to these tips. To date, the highest magnetic field gradients have been reported in studies of magnetic disk drive heads, ranging between $2 \times 10^7$ and $4 \times 10^7$ T/m [Tsang2006]. The pole tips used in drive heads are typically made of soft, high-moment materials like FeCo, and have widths below 100 nm. The FeCo magnetic tip used in the TMV experiment, on the other hand, was more than 200 nm in diameter, and generated a field gradient of $4 \times 10^6$ T/m. Moreover, calculations indicate that these tips did not achieve the ideal gradients which one would calculate assuming that they were made of pure magnetic material. This discrepancy may be due to a dead layer on the outside of the tips, to defects inside the tips, or to contamination of the magnetic material.

In 2012, Mamin et al. demonstrated the use of dysprosium (Dy) magnetic tips for MRFM [Mamin2012]. Dy has a bulk saturation magnetization up to 3.7 T compared to 2.4 T for the FeCo alloy, which was previously used for MRFM tips. Under similar experimental conditions (i.e. tip-sample spacing, temperature, and external magnetic field), the Dy tips produced $6 \times 10^6$ T/m, representing a modest improvement of 50%. For small spin ensembles, where the statistical polarization dominates [Sleator1985], the signal consists of the variance of the force, which implies that the required averaging time goes inversely with the fourth power of the field gradient [Mamin2003]. For this reason, even modest enhancements of the field gradient can be well worth the effort.

Despite this progress, magnetic tips producing such high gradients had not yet been realized on the cantilever force sensor. Moving to the "magnet-on-cantilever" rather than the "sample-on-cantilever" geometry enables the study of a broad range of samples. Having to attach the sample to an ultrasensitive cantilever puts constraints on samples size and poses the problem of attachment. In the "magnet-on-cantilever" geometry, one simply approaches the cantilever sensor to the sample, as in standard SPM. Possible target samples could then include, e.g., delicate biological samples that need to be embedded in a thin film of water and flash frozen to preserve their native structure, working organic semiconductor devices, or any nanometer-scale samples spread on a surface. However, the practical micro-fabrication challenge of realizing a high-quality nano-magnet *and* a high-quality mechanical sensor on the same device has proven difficult to overcome. After much process development, in 2012, Longenecker demonstrated MRFM of $^1$H in a polystyrene film using a "magnet-on-cantilever" configuration achieving gradients around $5 \times 10^6$ T/m [Longenecker2012]. The gradients achieved exceeded previous "magnet-on-cantilever" devices by a factor of 8, which, in principle, would allow for sub-10-nm resolution $^1$H MRI of samples on a surface.

In 2016, Tao demonstrated that a commercial hard-disk write head could be used to generate 5 times higher gradients than the Dy tips in an MRFM-type apparatus [TaoNC2016]. Experiments on the diamagnetic and paramagnetic forces generated by the write head reveal a gradient of $2.8 \times 10^7$ T/m within 5 nm of the surface. Crucially the magnetic field generated by the write head and its gradient are switchable in about 1 ns. The combination of large field and rapid switching should allow the implementation of very fast spin manipulation techniques and potentially open the way for high-resolution force NMR spectroscopy on nanometer-scale samples by force-detected means. Further desirable features include high-vacuum compatibility, low power dissipation, and an extremely flat surface topography amenable to follow-up lithography.

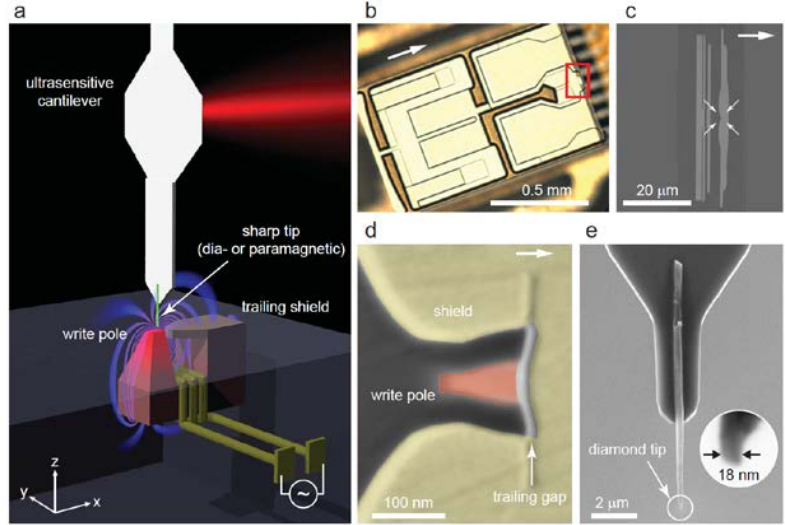

Figure 13: Geometry of write-head experiment. a, A sharp diamond needle (green), attached to a nano-mechanical force transducer, is positioned over the write pole of a magnetic recording head. An alternating current periodically switches the pole polarity and induces magnetic gradient forces through dia- or paramagnetism in the tip. Experiments are carried out in a SPM operating at 4 K and in high vacuum. b, Optical micrograph of a commercial write head. Arrows in b, c and d point in the direction of the trailing edge (in positive x direction). c, Zoom-in on the write/read region of the device. The write pole is at the center of the four arrows. d, The write pole (red) is surrounded by a return shield (yellow) that serves to recollect the field lines. The gradient is largest in the 20 nm-wide trailing gap between pole and shield. e, Diamond nanowire used to probe the local magnetic force. Inset shows apex of tip B [TaoNC2016].

Since the gradient strength falls off rapidly with distance, the ability to bring the sample to within 5 nm of the magnet without losing force sensitivity is crucial in Tao's realization of large $G$. Normally, measurements at small tip–sample spacing are hampered by strong tip–sample interactions which produce mechanical noise and dissipation in the cantilever. These interactions have been studied in similar systems [StipePRL2001C, KuehnPRL2006] and several mechanisms have been proposed to explain its origin depending on the details of the configuration [Persson1998, ZuritaSanchez2004, Volokitin2003, Volokitin2005, Labaziewicz2008]. Most explanations point to trapped charges or dielectric losses in either the substrate or the cantilever tip. Experimentally, several strategies can mitigate non-contact friction effects, including chemical modification of the surface, narrow tip size, or high-frequency operation. Tao and coworkers relied on a specially designed diamond NW tip producing exceptionally low non-contact friction [TaoNL2015]. The low dielectric constant, low loss tangent, and lack of defect-rich surface oxide make diamond the ideal material for a low-friction tip. Furthermore, NW tip radii of 19 nm with apex angles around 15˚ minimized the tip-surface interaction area.

### 8.2 MECHANICAL TRANSDUCERS

The second means to improving the signal-to-noise ratio is the development of more sensitive mechanical transducers, i.e. transducers that exhibit a lower force noise spectral density $S_F$. For a mechanical resonator, $S_F$ is given by:

$$S_F = 4\, k_B T\, \Gamma, \tag{5}$$

where $k_B$ is the Boltzmann constant, $T$ is the temperature, and $\Gamma$ is the resonator's mechanical dissipation. This term is related to the mechanical system's energy loss to the environment: $\frac{dE}{dt} = -\Gamma \dot{x}^2$, where $\dot{x}$ is the velocity of the resonator's displacement. The minimization of $S_F$ therefore involves reducing the operating temperature and the dissipation, which can also be written $\Gamma = \frac{m \omega_0}{Q}$, where $m$ is the motional mass of the mechanical resonator.

In practice, this means that at a given temperature, a well-designed cantilever must simultaneously have low $m\omega_0$ and large $Q$. For long and thin cantilevers, the Euler-Bernoulli beam equations imply that $m \omega_0 \propto \frac{w t^2}{l}$, while experiments show that $Q$ is limited by surface-related losses, as shown in Fig. 14. This effect leads to a linear decrease with increasing surface-to-volume ratio meaning that $Q \propto t$ [TaoNano2015]. Therefore, $\Gamma \propto \frac{w t}{l}$, meaning that long, narrow, and thin cantilevers are the most sensitive transducers. In fact, a review of real transducers confirms this trend. The ultimate force resolution of such devices, which inevitably have large surface-to-volume ratios, is limited by surface imperfections. For devices with extremely high surface-to-volume ratios, the reduction of $Q$ caused by these effects begins to compensate for gains made in $m\omega_0$.

Efforts in producing mechanical transducers with low dissipation can largely be divided into "top-down" and "bottom-up" approaches. The world's most sensitive transducers and some of its most common – including cantilevers used in AFM – are all fabricated using top-down methods. Currently, typical transducer fabrication processes involve optical or electron-beam lithography, chemical or plasma etching, and a release step. Even smaller structures can be milled out using focused ion beam techniques. New developments in bottom-up growth, however, are changing the status quo. Researchers can now grow nanometer-scale structures such as carbon nanotubes (CNTs) and nanowires (NWs) from the bottom-up with unprecedented mechanical properties. Unlike traditional cantilevers and other top-down structures, which are etched or milled out of a larger block of material, bottom-up structures are assembled unit-by-unit to be almost defect-free on the atomic-scale with perfectly terminated surfaces. This near perfection gives bottom-up structures a much smaller mechanical dissipation than their top-down counterparts while their high resonance frequencies allow them to couple less strongly to common sources of noise.

On the top-down side, Tao demonstrated the fabrication of ultra-sensitive cantilever made from single-crystal diamond with thickness down to 85 nm and quality factors exceeding 1 million at room temperature [Tao2014]. The corresponding thermal force noise at millikelvin temperatures for the best cantilevers was around 500 zN/Hz$^{1/2}$. This value represents a factor of 2 improvement on the Si cantilevers used in the TMV imaging experiment [Degen2009]. Despite the modest gain, the article shows the promise of using diamond material for ultrasensitive cantilevers, as shown in Fig. 14. Correcting for factors dependent on geometry, the authors show that diamond consistently outperforms Si in terms of a material for low mechanical dissipation resonators. The authors estimate that given observed trends and processing capabilities, diamond cantilevers with thicknesses of 50 nm, could be realized with low-temperature force sensitivities down to around 50 zN/Hz$^{1/2}$. Nevertheless, processing nanomechanical structures from diamond is far more expensive and difficult than from Si.

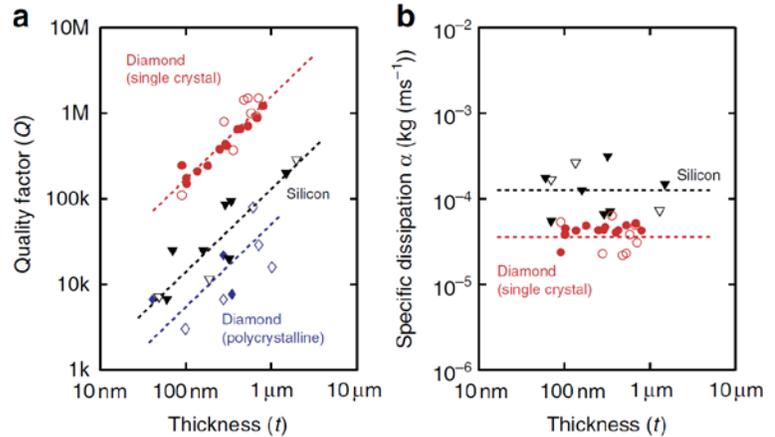

Figure 14: Comparison of Q between nanomechanical resonators made from different materials. (a) Comparison of Q highlighting that for similar device dimensions, Q of single-crystal diamond are consistently higher by about an order of magnitude over single-crystal silicon devices. (b) Comparison of the geometry-independent dissipation parameter. Open symbols are 300 K values and filled symbols are B4 K values. Dashed lines indicate linear thickness dependence of Q [TaoNatComm].

In a separate paper, Tao tackled the surface dissipation problem on Si cantilevers by attempting to modify and passivate the surface in an effort to produce more sensitive force transducers [TaoNano2015]. They found that the 1–2 nm-thick native oxide layer of silicon contributes to about 85% of the dissipation of the mechanical resonance. Through careful study, they observed that mechanical dissipation is proportional to the thickness of the oxide layer and that it crucially depends on oxide formation conditions. They further demonstrated that chemical surface protection by nitridation, liquid-phase hydrosilylation, or gas-phase hydrosilylation can inhibit rapid oxide formation in air and results in a permanent improvement of the mechanical quality factor between three- and five-fold. This improvement extends to cryogenic temperatures. Their results showed, that integrating the correct recipes with standard cleanroom fabrication can be extremely beneficial for ultrasensitive nanomechanical force sensors, including silicon cantilevers, membranes, and NWs.

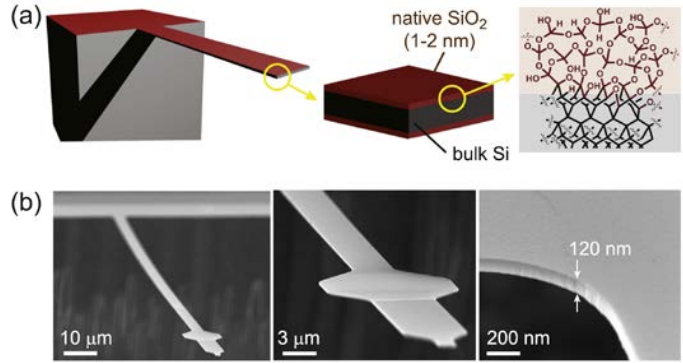

Figure 15: (a) Schematic buildup of silicon cantilever devices, showing the trimorph-like SiO$_2$(1 nm)–Si(120 nm)–SiO$_2$(1 nm) cross section and the atomistic makeup of the native surface oxide layer. Si atomic labels are omitted for clarity. (b) Scanning electron micrographs of one of the 120 nm thick cantilever devices used in this study [TaoNano2015].

On the bottom-up side, remarkable progress is being made. In two separate letters, Moser et al. demonstrated the use of a CNT as a sensitive force sensor with a thermally limited force sensitivity of 12 zN/Hz$^{1/2}$ at 1.2 K in 2013 [Moser2013] and then of 1 zN/Hz$^{1/2}$ at 44 mK in 2014 [Moser2014]. If such devices could be integrated into an MRFM set-up without degrading force sensitivity, detection of a single nuclear spin would be feasible. Nevertheless, there are factors that complicate the application of CNTs for force microscopy, including their very small linear dynamic range [Eichler2011] and the fact that their doubly clamped geometry is not easily amenable the protruding tip-like geometry of most SPM force sensors.

NW oscillators, on the other hand, have a large linear dynamic range, can be grown to many different sizes, and are more versatile and controllable than CNTs. Diameters range from tens to hundreds of nanometers and lengths reaching up to tens of microns. NWs can be grown from several materials including Si, GaAs, GaP, InAs, InP, GaN, and AlN. The central challenge facing NW mechanical sensors is the difficulty of detecting their displacement. A mechanical oscillator such as a cantilever or membrane is merely a transducer, i.e. an element which transforms a force into a displacement. For a force to be measured, the resulting displacement must be detected. Various techniques exist to detect the displacement of traditional micromechanical oscillators including optical, microwave, capacitive, magnetic, and piezoelectric schemes. The sensitivity of convenient optical techniques such as beam deflection or interferometry suffers as the dimensions of the mechanical resonator become smaller than the wavelength of light.

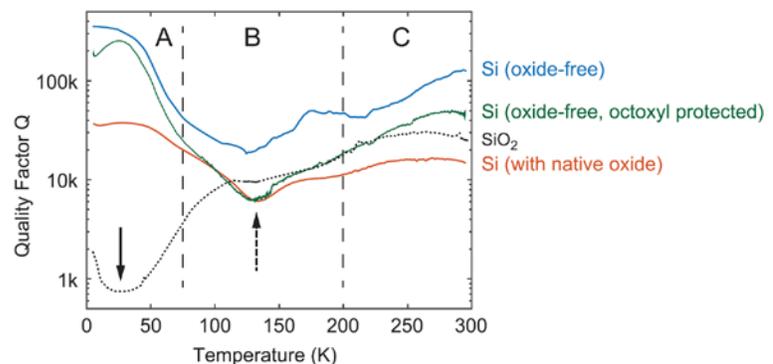

Figure 16: $Q$ as a function of temperature for three Si cantilevers (oxide-free, octoyl-protected, native oxide) and a SiO$_2$ cantilever. A, B, and C roughly divide between regions with different sources of friction. Solid arrow indicates dissipation peak caused by SiO$_2$, and dashed arrow indicates dissipation peak possibly caused by the organic protection layer and surface [TaoNano2015].

In 2008, however, Nichol et al. demonstrated a polarization-enhanced interferometry technique capable of detecting the thermal motion of a Si NW with a diameter less than 100 nm [Nichol2008]. A more detailed study of the limits of optical detection was carried out by Ramos et al. in 2013, finding that displacement sensitivities of 1 fm/Hz$^{1/2}$ can be achieved for 50-nm-diameter NWs [Ramos2013]. Once the thermal motion of a mechanical transducer can be measured, the combined system is a thermally limited force sensor – a system whose minimum detectable force is solely determined by its thermal fluctuations.

Nichol et al. went further in a subsequent 2012 paper and used their Si nanowire force transducers in an MRFM experiment detecting $^1$H in a nanometer-scale polystyrene sample [Nichol2012]. During the measurements they achieved a thermally limited force sensitivity of around 1 aN/Hz$^{1/2}$ at a spacing of 80 nm from the surface at 8 K, which is significantly lower than was measured at 300 mK in the TMV experiment [Degen2009]. This improvement is largely due to the ultra-low native dissipation of the NWs in comparison to top-down ultrasensitive cantilever and to their drastically reduced surface dissipation. In fact, Nichol et al., show that at a tip-surface spacing of 7 nm, a typical Si NW experiences nearly a factor of 80 less surface dissipation and factor of 250 less total dissipation than audio frequency cantilevers under similar conditions. The mechanisms behind this difference are not completely clear; the small cross-sectional area of a NW may decreases its coupling to the surface or, perhaps, the spectral density of surface fluctuations is lower at the MHz resonant frequencies of the NWs that at the kHz resonant frequencies of the cantilevers.

This ground-breaking work established NW oscillators as ultrasensitive cantilevers for MRFM detection. As discussed in a later section, the measurement protocol that was developed for the NW transducers uses a nanoscale current-carrying wire to generate both time-dependent RF magnetic fields and time dependent magnetic field gradients. This protocol, known as MAGGIC, ultimately opened new avenues for nanoscale magnetic resonance imaging with more favorable SNR properties [NicholPRX2013].

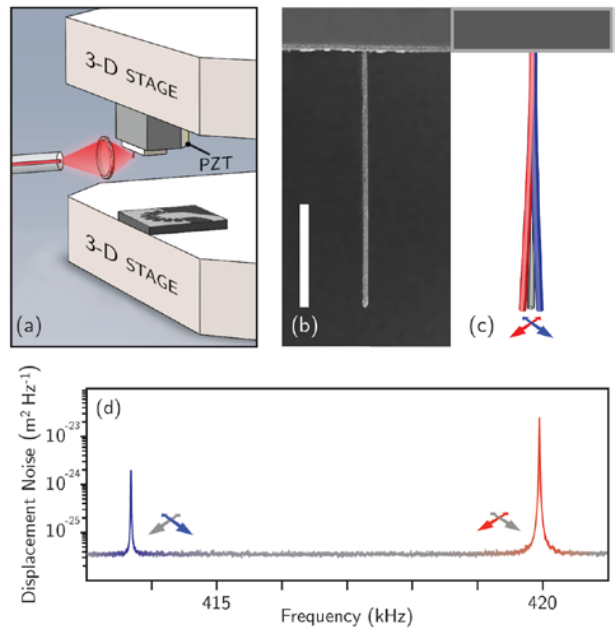

Figure 17: NW force sensors. (a) A fiber-optic interferometer is aligned with a single NW using a piezoelectric positioning stage (top). A second stage (bottom) is used to position and scan the sample surface under the NW. (b) A SEM of a GaAs/AlGaAs NW. The scale bar represents 10 μm. (c) A schematic diagram showing the two orthogonal fundamental flexural modes of the NW. (d) The displacement spectral noise density of the fundamental mode doublet measured by fiber-optic interferometry [Rossi2016].

A NW's highly symmetric cross-section results in orthogonal flexural mode doublets that are nearly degenerate [Nichol2008, Li2008]. This property makes bottom-up grown NWs extremely sensitive vectorial force sensors. In the pendulum geometry, these modes can be used for the simultaneous detection of in-plane forces and spatial force derivatives along two orthogonal directions [Gloppe2014]. Although one-dimensional (1D) dynamic lateral force microscopy can be realized using the torsional mode of conventional AFM cantilevers [Pfeiffer2002, Giessibl2002, Kawai2005, Kawai2009, Kawai2010], the ability to simultaneously image all vectorial components of nanoscale force fields is of great interest. Not only would it provide more information on tip-sample interactions, but it would also enable the investigation of inherently 2D effects, such as the anisotropy or non-conservative character of specific interaction forces.

Two 2017 experiments have recently extended the application of these vectorial force sensors. Mercier de Lépinay et al. have used a NW to map the electrostatic forces of a charged tip [Mercier2017], while Rossi et al. have taken advantage of the NW's adaptability as a scanning probe to image a sample surface [Rossi2017]. In the latter work, the authors show that this universally applicable technique enables a form of AFM particularly suited to mapping the size and direction of weak tip-sample forces. The potential of using this vectorial sensitivity in a purpose built MRFM apparatus is also excellent.

### 8.3 MEASUREMENT PROTOCOLS

In addition to improvements in MRFM hardware, the last few years have also yielded a variety of promising new measurement schemes promising to improve SNR and therefore reduce measurement times.

In 2010, Oosterkamp et al. demonstrated the detection of multiple MRFM signals simultaneously, both from different nuclear species and distinct sample positions using frequency domain multiplexing [Oosterkamp2010]. The protocol took advantage of the wider effective noise bandwidth of the damped cantilever transducer compared with the NMR signal bandwidth. A similar signal multiplexing technique was demonstrated by Moores in 2015, where the signals from different nuclear spin ensembles are encoded in the phase of the cantilever force signal. In this experiment statistically polarized spin signals from two different nuclear species and six spatial locations were collected simultaneously leading to a one-dimensional imaging resolution better than 5 nm [Moores2015]. Applied together, these results allow – in principle – for reductions in integration times over 10-fold.

In 2011, Xue et al. introduced a slight variation on the standard MRFM geometry where the long axis of the cantilever is normal to both the external magnetic field and the RF microwire source [XueAPL2011]. This configuration avoids any magnetic field induced mechanical dissipation in the cantilever, which generally imposed practical limitations on the applied external field or the measurement sensitivity. The same year in a second paper, Xue et al. measured MRFM signal from nuclear spin in a nanometer-scale semiconductor sample [XuePRB2011]. The work provided a detailed analysis of the MRFM receptivity of quadrupolar nuclei for both Boltzmann polarized and statistically polarized ensembles. The authors found that MRFM receptivity scales more favorably than conventional receptivity for low-γ nuclei such as those found in GaAs and other semiconductors. These results are particularly promising for efforts aimed at using MRFM for subsurface, isotopically selective imaging on nanometer-scale III-V samples, especially since conventional methods such as SEM and TEM lack isotopic contrast.

The nanometer-scale spin ensembles typically measured by MRFM differ from larger ensembles in that random fluctuations in the total polarization— also known as spin noise—exceed the normally dominant mean thermal polarization. This characteristic imposes important differences between nanoMRI and conventional MRI protocols. In the former technique, statistical fluctuations are usually measured, whereas in the latter the signal is based on the thermal polarization [Bloch1946, Degen2007, Peddibhotla2013]. The thermal polarization— also known as Boltzmann polarization—results from the alignment of nuclear magnetization under thermal equilibrium along a magnetic field. The statistical polarization, on the other hand, arises from the incomplete cancellation of magnetic moments within the ensemble.

In order to compare the thermal and the statistical polarization, we express both as fractions of a fully polarized system $M_{100\%} = N\hbar\gamma I$, resulting in $P_{thermal} = \frac{M_z}{M_{100\%}} = \frac{I+1}{3}\frac{\hbar\gamma B}{k_B T}$ and $P_{statistical} = \frac{\sigma_{M_z}}{M_{100\%}} = \sqrt{\frac{I+1}{3I}\frac{1}{N}}$, where $N$ is the number of spins in the detection volume, $\hbar$ is Planck's constant, $\gamma$ is the gyromagnetic ratio, $I$ is the spin number, $k_B$ is Boltzmann's constant, and $T$ is the temperature. Note that while $P_{thermal}$ is independent of the ensemble size, $P_{statistical}$ increases with decreasing ensemble size. This implies that for ensembles with $N < N_c$, where $N_c$ is some critical number of spins reflecting the border of the two regimes, $P_{statistical} > P_{thermal}$. For this ensemble size, the size of the natural spin polarization fluctuations will begin to exceed the magnitude of the mean polarization in thermal equilibrium. This transition typically occurs on the micro- or nanometer-scale, underpinning the dominant role that statistical fluctuations play in nanometer-scale NMR. Furthermore, by measuring both mean thermal magnetization and the standard deviation, one can determine the number of spins in the detected ensemble depending on the ratio of $M_z$ and $\sigma_{M_z}$:

$$N = \frac{3}{I(I+1)} \left(\frac{k_B T}{\hbar \gamma B}\right)^2 \left(\frac{M_z}{\sigma_{M_z}}\right)^2. \qquad (5)$$

Note that for $M_z = \sigma_{M_z}$, the ensemble contains $N = N_c$ spins. In a material with a nuclear spin density $na$, where $n$ is the number density of the nuclear element and $a$ is the natural abundance of the measured isotope, the corresponding detection volume is then given by $V = \frac{N}{na}$. This transition from a thermally dominated to a statistically dominated ensemble magnetization and a scheme for determining the number of spins in a nanometer-scale ensemble was explicitly demonstrated by Herzog et al. in 2014 [Herzog2014].

A number of protocols in recent years have been developed specifically for working with statistically polarized nuclear spins. In 2008, IBM scientists were able – for the first time – to follow the fluctuations of a statistical polarization of nuclear spins in real time. These experiments followed the dynamics of an ensemble of roughly 2 × 10⁶ ¹⁹F spins in CaF$_2$ [Degen2007]. The challenge of measuring statistical fluctuations presents a major obstacle to nanoscale imaging experiments. In particular, the statistical polarization has a random sign and a fluctuating magnitude, making it hard to average signals. An efficient strategy for imaging spin fluctuations is therefore to use polarization variance, rather than the polarization itself, as the image signal. This was demonstrated both by force-detected [Degen2009, Mamin2007, Mamin2009, Degen2007] and conventional [Muller2006] MRI. Furthermore, it was demonstrated that for cases where spin lifetimes are long, rapid randomization of the spins by RF pulses can considerably enhance the signal-to-noise ratio of the image [Degen2007]. In the end, for the purposes of imaging, it is not necessary to follow the sign of the spin polarization; it is enough to simply determine from the measured spin noise how many spins are present at a particular location.

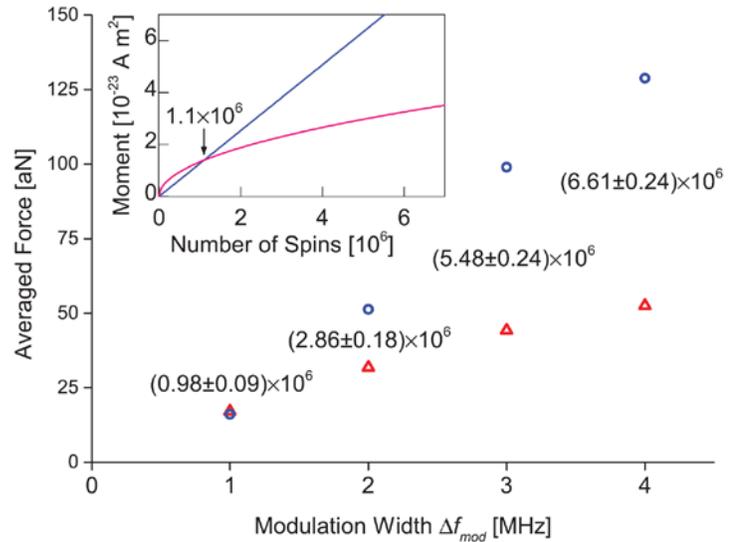

Figure 18: Mean force (blue circles), originating from the thermal polarization, and standard deviation (red triangles), originating from the statistical polarization, as a function of pulse modulation width at B=4.37 T and T=4.4 K. The values between the symbols show the corresponding number of spins N. Inset: A theoretical plot for ¹⁹F as a function of N showing the crossover at N$_c$. The similarity between the inset and the figure indicate that the number of detected spins (or the detection volume V) is roughly proportional to modulation width [Herzog2014].

The nuclear spin lifetime itself, which is apparent as the correlation time of the nuclear fluctuations $\tau_m$, was also shown to be an important source of information. Using suitable RF pulses, researchers demonstrated that Rabi nutations, rotating-frame relaxation times, and nuclear cross polarization can be encoded in $\tau_m$ leading to new forms of image contrast [Poggio2007, Poggio2009]. In 2009, the IBM group exploited couplings between different spin species to enhance the 3D imaging capability of MRFM with the chemical selectivity intrinsic to magnetic resonance. They developed a method of nuclear double resonance that allows the enhancement of the polarization fluctuation rate of one spin species by applying an RF field to the second spin species, resulting in suppression of the MRFM signal [Poggio2009]. The physics behind this approach is analogous to Hartmann–Hahn cross polarization (CP) in NMR spectroscopy [Hartmann1962], but involves statistical rather than Boltzmann polarization. The IBM group was inspired by previous work done with Boltzmann polarizations at the ETH in Zürich demonstrating CP as an efficient chemical contrast mechanism for micrometer-scale one-dimensional MRFM imaging [Lin2006, Eberhardt2007, Eberhardt2008]. In the IBM experiment, MRFM was used to measure the transfer between statistical polarizations of $^1$H and $^{13}$C spins in $^{13}$C-enriched stearic acid. The development of a cross-polarization technique for statistical ensembles adds an important tool for generating chemical contrast to the recently demonstrated technique of nanometer-scale MRI.

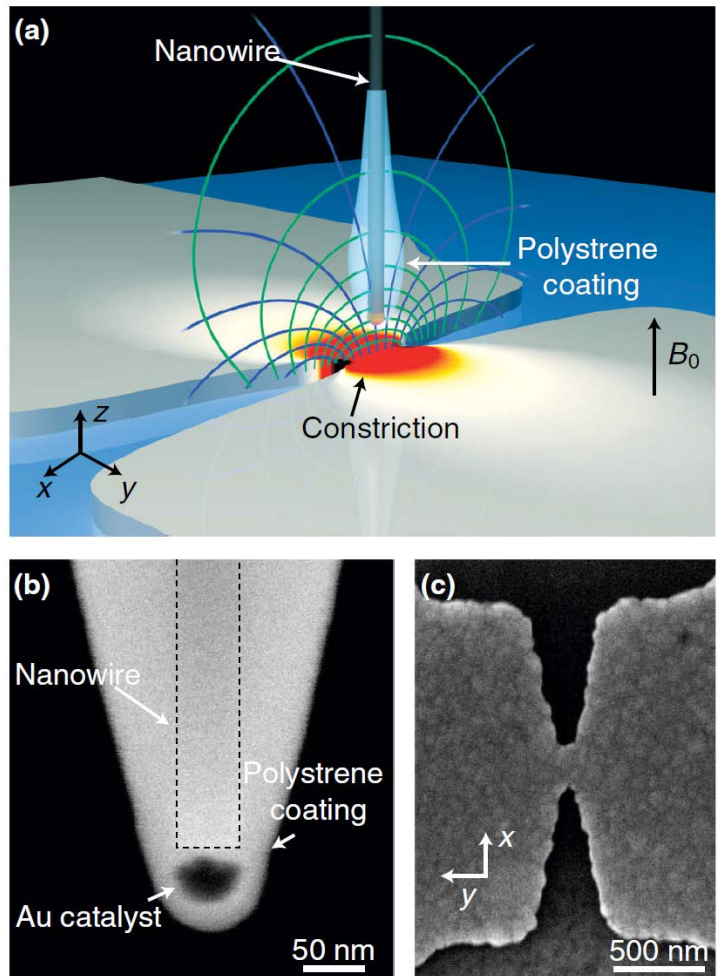

Figure 19: (a) Schematic of the experimental setup. A silicon nanowire coated with polystyrene is positioned near the constriction in a Ag current-carrying wire. The locally high current density through the constriction generates intense fields and gradients used for readout, spin manipulation, and spatial encoding. During imaging, the spin density is encoded along contours of constant Larmor and Rabi frequencies, which are illustrated as blue and green lines, respectively. (b) SEM of a representative nanowire and polystyrene coating prepared in the same manner as the nanowire and sample used in this study. The dashed lines indicate the outer diameter of the nanowire. (c) Scanning electron micrograph of the constriction used in this study [NicholPRX].

In 2013, Peddibhotla et al. demonstrated a technique to create spin order in nanometer-scale ensembles of nuclear spins by harnessing these fluctuations to produce polarizations both larger and narrower than the thermal distribution [Peddibhotla2013]. Although the results were obtained with a low-temperature MRFM, the capture and storage of spin fluctuations is generally applicable to any technique capable of detecting and addressing nanometer-scale volumes of nuclear spins in real time. When polarization cannot be created through standard hyperpolarization techniques such as dynamic nuclear polarization, this method provides a viable alternative. One could imagine, for instance, such nuclear polarization capture processes enhancing the weak MRI signals of a nanometre-scale $^1$H-containing biological sample or of a semiconducting nanostructure.

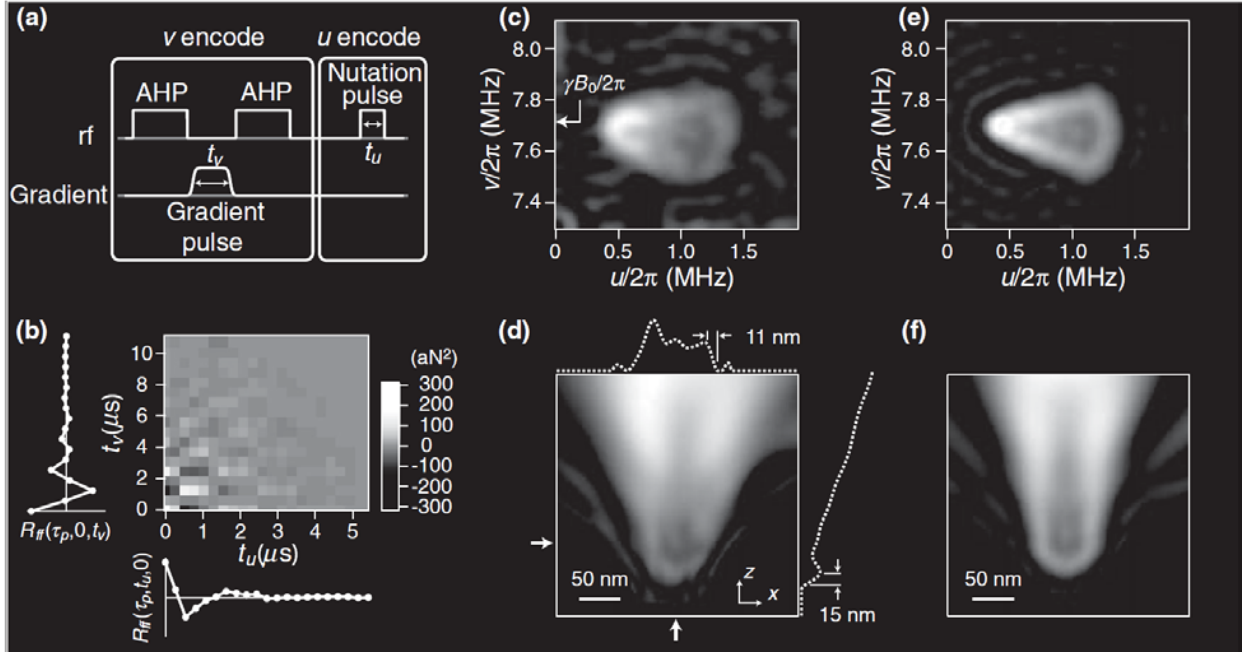

Figure 20: Two-dimensional MRI of the polystyrene sample. (a) Image encoding sequence. (b) Raw data. (c) Signal density in the ðu; vÞ coordinate system obtained by cosine transforming the raw data. (d) Real-space reconstruction of the projected spin density. The nanowire and gold catalyst are clearly visible through the polystyrene in the image as a reduction in the spin density. The cross sections above and to the right of the image are taken along the lines indicated by the arrows. (e) Simulated signal density. (f) Real-space reconstruction of the simulation in (e) [NicholPRX2013].

In 2016, Issac et al. tested a method designed to circumvent MRFM's reliance on weak statistical spin polarizations [Issac2016]. The authors applied dynamic nuclear polarization (DNP), which relies on the transfer of magnetization from electron spins to nuclear spins in a sample, to enhance the mean magnetization of the MRFM detection volume. In particular, the experiment applied the widely applicable cross-effect DNP mechanism to create hyper-thermal nuclear spin polarization in a thin-film polymer sample in a "magnet-on-cantilever" MRFM experiment. As discussed in the article, although a number of challenges still need to be addressed, using DNP to create hyper-thermal spin polarization in an MRFM experiment offers many exciting possibilities for increasing the technique's imaging sensitivity.

### 8.4 NANOMRI WITH A NANOWIRE FORCE SENSOR

Perhaps the single most promising result since the TMV imaging experiment was demonstrated by Nichol et al. in 2013 [NicholPRX2013]. The authors report on a modified MRFM imaging protocol obtaining a 2D projection of $^1$H density in a polystyrene sample with approximately 10-nm resolution. The measurement, which relied on statistical polarization for signal contrast, used a bottom-up Si NW mechanical oscillator as the force transducer. Furthermore, the authors used a nanometer-scale metallic constriction to produce both the RF field and a switchable magnetic field gradient.

Given that nanometer-scale MRFM requires intense static magnetic field gradients, both NMR spectroscopy and uniform spin manipulation using RF pulses have always been difficult to implement in such measurements. In addition, conventional pulsed magnetic resonance techniques cannot be applied to nanometer-scale MRFM because statistical spin fluctuations often exceed the Boltzmann spin polarization. In this regime, the projection of the sample magnetization along any axis fluctuates randomly in time.

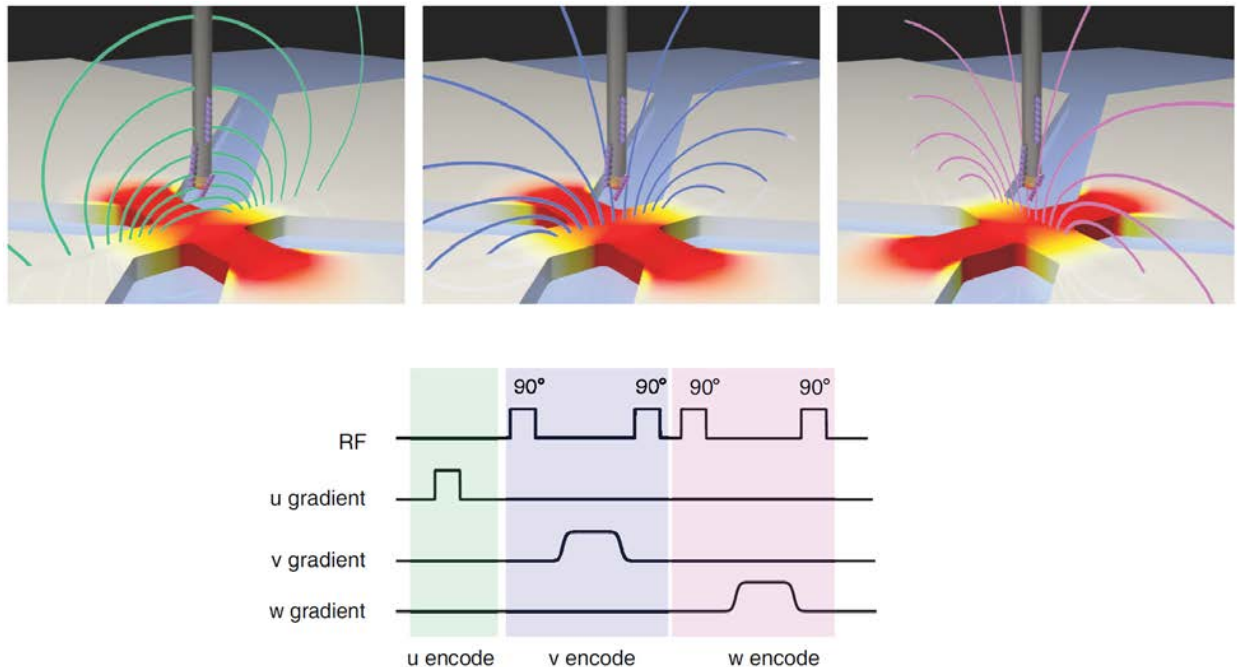

Figure 21: Illustration of a potential 3D encoding sequence. An RF gradient pulse and two successive orthogonal static gradients could be applied to perform three-dimensional Fourier transform imaging. Samples in future experiments are likely to be biolmolecules attached to the Si NW tip [NicholPhD2013].

In their article, Nichol et al. presented a new paradigm in force-detected magnetic resonance that overcomes both challenges to enable pulsed nuclear magnetic resonance in nanometer-size statistically polarized samples. The first challenge was solved by using the nanometer-scale constriction to generate both large RF fields and large magnetic field gradients. In this way, the authors were able to turn their magnetic field gradients and on and off at will. Using a scheme similar to conventional MRI, using switchable gradients in static and RF field, they encoded the Fourier transform of the 2D spin density into their spin signal. As a result, they were able to reconstruct a 2D projection of the $^1$H density in a polystyrene sample with roughly 10-nm resolution. The protocol was able to function in the statistically polarized regime because the authors periodically applied RF pulses, which create correlations in the statistical polarization of a solid organic sample. The spin-noise correlations were then read-out using gradient pulses generated by ultra-high current densities in the nanoscale metal constriction. The authors also showed that Fourier-transform imaging enhances sensitivity via the multiplex advantage for high-resolution imaging of statistically polarized samples. Most importantly, the protocol established a method by which all other pulsed magnetic resonance techniques can be used for nanoscale imaging and spectroscopy.

The authors' work is ground-breaking on several levels. From a technical point of view, they showed how a bottom-up NW can be successfully used as a force sensor for nano-MRI. Given the potential for even more sensitive NW transducers, this proof-of-concept experiment bodes well for increasing nano-MRI sensitivity and resolution. Even without improvement in sensitivity, the authors' technique could also be extended to enable full 3D encoding with constrictions capable of producing two orthogonal static gradients, as shown in Fig. 21 [NicholPhD2013]. More generally, the approach serves as a model for applying sophisticated pulsed magnetic resonance schemes from conventional MRI to the nanometer-scale version.

## 9 COMPARISON TO OTHER TECHNIQUES

The unique position of MRFM among high-resolution microscopies becomes apparent when comparing it to other, more established nanoscale imaging techniques. As a genuine scanning probe method, MRFM has the potential to image matter with atomic resolution. While atomic-scale imaging is routinely achieved in scanning tunneling microscopy and atomic force microscopy, these techniques are confined to the top layer of atoms and cannot penetrate below surfaces [Hansma1987, Giessibl2003]. Moreover, in standard scanning probe microscopy (SPM), it is difficult and in many situations impossible to identify the chemical species being imaged. Since MRFM combines SPM and MRI, these restrictions are lifted. The three-dimensional nature of MRI permits acquisition of sub-surface images with high spatial resolution even if the probe is relatively far away. As with other magnetic resonance techniques, MRFM comes with intrinsic elemental contrast and can draw from established NMR spectroscopy procedures to perform detailed chemical analysis. In addition, MRI does not cause any radiation damage to samples, as do electron and x-ray microscopies.

MRFM also distinguishes itself from super-resolution optical microscopies that rely on fluorescence imaging [Huang2009]. On the one side, optical methods have the advantage of working in vivo and they have the ability to selectively target the desired parts of a cell. Fluorescent labeling is now a mature technique which is routinely used for cellular imaging. On the other side, pushing the resolution into the nanometer range is hampered by fundamental limitations, in particular the high optical powers required and the stability of the fluorophores. Moreover, fluorescent labeling is inextricably linked with a modification of the target biomolecules, which alters the bio-functionality and limits imaging resolution to the physical size of the fluorophores.

MRFM is also unique among other nanoscale spin-detection approaches. While single electron spin detection in solids has been shown using several techniques, these mostly rely on the indirect read-out via electronic charge [Elzeran2004, Xiao2004] or optical transitions [Wrachtrup1993, Jelezko2002]. In another approach, the magnetic orientation of single atoms has been measured via the spin-polarized current of a magnetic STM tip or using magnetic exchange force microscopy [Durkan2002, Heinze2000, Kaiser2007]. These tools are very valuable to study single surface atoms, however, they are ill-suited to map out sub-surface spins such as paramagnetic defects. In contrast, MRFM directly measures the magnetic moment of a spin, without resorting to other degrees of freedom, making it a very general method. This direct measurement of magnetic moment (or magnetic stray field) has also been carried out on the nanometer-scale using other techniques including, Hall microscopy [Chang1992], SQUID microscopy [Kirtley1995, Vasyukov2013], or magnetometry based on single nitrogen-vacancy (NV) centers in diamond [Degen2008, Maze3008, Balasubramanian2008]. So far, only NV magnetometry has been used in combination with NMR, demonstrating the capability for nano MRI.

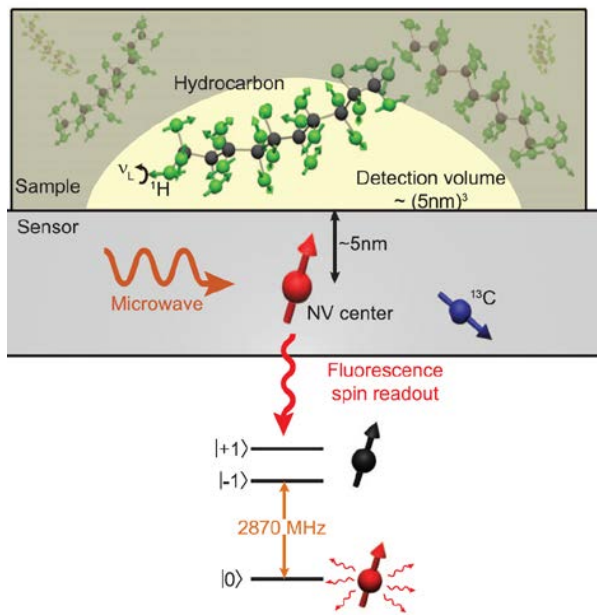

Figure 22: NV centers implanted near the diamond surface were used to detect $^1$H spins within liquid and solid organic samples placed on the crystal surface [Staudacher2013].

The NV center is a defect center in crystalline diamond consisting of a nitrogen atom adjacent to a vacancy in the lattice. This complex acts as a single spin-1 defect. The quantum state of the NV center can be initialized and read-out using a visible light and because of its long coherence time has been used as a sensor of fluctuating magnetic fields with a sensitivity down to 10 nT/Hz$^{1/2}$. High sensitivity is maintained even under ambient conditions, making the technique extremely promising for in vivo nanoMRI [Bhallamudi2015]. MRFM, on the other hand, requires high-vacuum and low temperature in order to reduce the thermal motion of the mechanical sensor and achieve high spin sensitivity.

The high sensitivity of the NV center, however, is only realized with defects that are deep enough within the diamond lattice to maintain long coherence times – typically deeper than 5 nm from the surface. This limitation has a strong consequence on sensing applications given that the sensitivity of the NV to magnetic moments, such as nuclear spins, depends on the dipole-dipole interaction. This interaction drops off as $r^{-3}$, where $r$ is the separation between the NV and the target magnetic moment. For this reason, detecting nuclear spins requires making trade-offs between using shallow enough NV centers such that they are strongly coupled to external nuclear spins, but deep enough that their coherence times do not drastically limit the moment sensitivity.

Despite these demanding requirements, impressive and rapid progress has been made in detecting nuclear magnetization using NV sensors. In 2013, two groups, Mamin et al. and Staudacher et al., reported NMR from (5 nm)$^3$ volumes of $^1$H spins on a diamond surface [Mamin2013, Staudacher2013]. In 2014, Loretz et al. decrease the detection volume to (1.8 nm)$^3$, which corresponds to 330 $^1$H spins [Loretz2014]. While sensitivity to a single $^1$H spin external to the diamond lattice still has to be unambiguously demonstrated [Loretz2015], it now appears that reaching this milestone is simply a matter of time. In 2015, three research group made the first step toward nuclear MRI with NV centers [Rugar2015, Haberle2015, DeVience2015]. In one case, Rugar et al. produce 2D image of $^1$H NMR from a polymer test sample using a single NV center sensor [Rugar2015]. As the sample was scanned past the NV center, it was used to detect the oscillating magnetic field from the sample's precessing $^1$H spins. The experiment achieved a spatial resolution of just over 10 nm. This work, as well as the two others, showed that 2D nanoMRI can be achieved using the simple concept of scanning an organic sample past a near-surface NV center [Bhallamudi2015].

There remains significant room for improvement in the technique. For example, the coupling of the NV center to the nuclear moments in the sample can be increased by using shallower NVs or by using improved detection protocols, such as double quantum magnetometry [Mamin2014]. MRI can – in principle – be extended to 3D with greatly enhanced spatial resolution by introducing a sufficiently strong source of magnetic field gradients, such as a small ferromagnet [Grinolds2014, Mamin2012].

Nevertheless, 3D imaging has not yet been demonstrated and it is unclear whether the demanding sample preparation can be overcome. For example, the ubiquitous $^1$H contamination layer on all samples is a serious obstacle to any molecular structure imaging applications since it is an interfering NMR signal source. Furthermore, even a shallow NV center with a long coherence time is only capable of detecting local magnetic fields, thus prohibiting depth resolution beyond a few nanometers. This fundamental "near-sightedness" of NV magnetometry puts a limit on the sample size that can be investigated. As a result, NV magnetometry cannot be used for 3D MRI imaging objects on the scale of 100 nm with a resolution of less the 5 nm.

This length-scale is referred to as the "unbridged regime" [Subramaniam2005, Moores2016] and happens to be a major blind spot for all known 3D imaging techniques. For this reason, there are many classes of structures that cannot be imaged, creating a blind spot for structural biologists. Several methods are under intense research and development to resolve objects within this regime in 3D, including super-resolution microscopy, cryo-electron microscopy, and nanoMRI. MRFM measures magnetic moment rather than magnetic field and it derives its resolution from the size of the magnetic field gradient and the sensitivity of its mechanical sensor. Therefore, it does not suffer from the same "near-sightedness" that NV magnetometry does. For this reason, it is still the ideal technique with which to tackle the challenges of the unbridged regime.

## 10 OUTLOOK

Although MRFM researchers have not surpassed the sub-5-nm nanoMRI resolution demonstrated in 2009, combining recent improvements in cantilever transducers, gradient sources, and detection protocol into a single apparatus should lead to 1-nm resolution nanoMRI with a range of roughly 100 nm. Further development of the paradigm shifting NW detection and Fourier encoding of the Illinois group may lead to more dramatic gains. Such progress would put the capabilities of nanoMRI by MRFM well into the "unbridged regime" and would distinguish it from NV-center magnetometry, which continues to be developed for short-range atomic-scale imaging. Nevertheless, several important obstacles must be overcome in order to turn the MRFM technique into a useful tool for biologists and materials scientists.

Most existing MRFM instruments are technically involved prototypes; major hardware simplifications will be required for routine screening of nanoscale samples. Suitable specimen preparation methods must be developed that are compatible with the low-temperature, high vacuum environment required for the microscope to operate at its highest sensitivity and resolution. While this is particularly challenging for biological samples, protocols exist which could be adapted to MRFM. In cryo-electron microscopy, for example, dispersed samples are vitrified to preserve their native structure by plunge-freezing in liquid nitrogen [Taylor1974]. As objects become smaller, isolation of samples and suppression of unwanted background signals from surrounding material will become increasingly important.

The conditions under which the latest MRFM imaging experiments were carried out are remarkably similar to those prevailing in cryo-electron microscopy, the highest-resolution 3D imaging technique commonly used by structural biologists today. Cryo-electron microscopy, like MRFM, operates at low temperatures and in high vacuum, requires long averaging times (on the order of days) to achieve sufficient contrast, and routinely achieves resolutions of a few nanometers [Lucic2005, Subramaniam2005]. Unlike MRFM, however, electron microscopy suffers from fundamental limitations that severely restrict its applicability. Specimen damage by high-energy electron radiation limits resolution to 5–10 nm if only a single copy of an object is available. Averaging over hundreds to thousands of copies is needed to achieve resolutions approaching 1 nm [Glaeser2008]. In addition, unstained images have intrinsically low contrast, whereas staining comes at the expense of modifying the native structure.

MRFM has the capability to image nanoscale objects in a non-invasive manner and to do so with intrinsic chemical selectivity. For this reason, the technique has the potential to extend microscopy to the large class of structures that show disorder and therefore cannot be averaged over many copies. These structures include such prominent examples as HIV, Influenza virus, and Amyloid fibrils. Virtually all of these complexes are associated with important biological functions ranging from a variety of diseases to the most basic tasks within the cellular machinery. For such complexes, MRFM has the potential not only to image the three-dimensional macromolecular arrangement, but also to selectively image specific domains in the interior through isotopic labeling.

While the most exciting prospect for MRFM remains its application to structural imaging in molecular biology, its applications are not limited to biological matter. For example, most semiconductors contain non-zero nuclear magnetic moments. Therefore, MRFM may prove useful for sub-surface imaging of nanoscale electronic devices. MRFM also appears to be the only technique capable of directly measuring the dynamics of the small ensembles of nuclear spin that limit electron spin coherence in single semiconductor quantum dots. Polymer films and self-assembled monolayers—important for future molecular electronics—are another exciting target for MRFM and its capability to image chemical composition on the nanoscale. Finally, isotopically engineered materials are becoming increasingly important for tuning a variety of physical properties such as transport and spin. Researchers currently lack a general method for non-invasively imaging the isotopic composition of these materials [Shimizu2006, Shlimak2001, Kelly2007]; MRFM techniques could fill this void.

## 11 CONCLUSION

Over the last 25 years, MRFM has led to exciting progress in the field of ultrasensitive spin-detection and high-resolution MRI microscopy. Starting with early demonstrations in the 1990s imaging with resolutions of a few micrometers—on par with conventional MRI microscopy—the technique has progressed to the point where it can resolve single virus particles and molecular monolayers. Recent improvements in various components have put 1-nm resolution within reach without major modifications to the instrument. Extremely promising new bottom-up transducers and the application of Fourier transform imaging techniques may provide even larger gains. Nevertheless, in addition to these improvements in the detetion hardware, much work still remains to be done in specimen preparation protocols, such that this resolution can be applied to 3D imaging of nano-biological samples or macromolecular complexes. The extension of MRFM to atomic resolution, where atoms in molecules could be directly mapped out and located in 3D, remains an exciting if technically very challenging prospect.

## 12 REFERENCES

[Akamine1990] Akamine et al., Appl. Phys. Lett. 57, 316 (1990).
[Alzetta1967] Alzetta et al., Il Nuovo Cimento B 62, 392 (1967).
[Ascoli1996] Ascoli et al., Appl. Phys. Lett. 69, 3920 (1996).
[Balasubramanian2008] Balasubramanian et al., Nature 455, 648 (2008).
[Barbic2009] Barbic, Magnetic Resonance Microscopy, ed Codd and Seymour, New York: Wiley, pp. 49-63 (2009).
[Berman2006] Berman et al., Magnetic Resonance Force Microscopy and a Single-Spin Measurement, Singapore: World Scientific (2006).
[Bhallamudi2015] Bhallamudi and Hammel, Nature Nanotech. 10, 104 (2015).
[Binnig1982] Binnig et al., Phys. Rev. Lett. 49, 57 (1982).
[BinnigPat1986] Binnig, Us Patent No. 4,724,318 (1986)
[BinnigPRL1986] Binnig et al., Phys. Rev. Lett. 56, 930 (1986).
[Blank2003] Blank et al., J. Magn. Reson. 165, 116 (2003).
[Bloch1946] Bloch, Phys. Rev. 70, 460 (1946).
[Bruland1998] Bruland et al., Appl. Phys. Lett. 73, 3159 (1998).
[Budakian2005] Budakian et al., Science 307, 408 (2005).
[Chang1992] Chang et al., Appl. Phys. Lett. 61, 1974 (1992).
[Chao2004] Chao et al., Rev. Sci. Instrum. 75, 1175 (2004).
[Chui2003] Chui et al., TRANSDUCERS, 12th Int. Conf. on Solid-State Sensors, Actuators and Microsystems vol 2 p 1120 (2003).
[Ciobanu2002] Ciobanu et al., J. Magn. Res. 158, 178 (2002).
[Degen2005] Degen et al., Phys. Rev. Lett. 94, 207601 (2005).
[Degen2007] Degen et al., Phys. Rev. Lett. 99, 250601 (2007).
[Degen2008] Degen, Appl. Phys. Lett. 92, 243111 (2008).
[Degen2009] Degen et al., Proc. Natl. Acad. Sci. U.S.A. 106, 1313 (2009).
[deLoubens2007] de Loubens et al., Phys. Rev. Lett. 98, 127601 (2007).
[DeVience2015] DeVience et al., Nature Nanotech. 10, 129 (2015).
[Dobigeon2009] Dobigeon et al., IEEE Trans. Image Process. 18, 2059 (2009).
[Durkan2002] Durkan and Welland, Appl. Phys. Lett. 80, 458 (2002).
[Eberhardt2007] Eberhardt et al., Phys. Rev. B 75, 184430 (2007).
[Eberhardt2008] Eberhardt et al., Angew. Chem. Int. Edn 47, 8961 (2008).
[Eichler2011] Eichler et al., Nature Nanotech. 6, 339 (2011).
[Elzerman2004] Elzerman et al., Nature 430, 431 (2004).
[Evans1956] Evans, Phil. Magn. 1, 370 (1956).
[Garner2004] Garner et al., Appl. Phys. Lett. 84, 5091 (2004).
[Giessibl2002] Proc. Natl. Acad. Sci. U.S.A. 99, 12006 (2002).
[Giessbl2003] Rev. Mod. Phys. 75, 949 (2003).
[Glaeser2008] Glaeser, Phys. Today 61, 48 (2008).
[Gloppe2014] Gloppe et al., Nature Nanotech. 11, 920 (2014).
[Grinolds2014] Grinolds et al., Nature Nanotech. 9, 279 (2014).
[Haberle2015] Häberle et al., Nature Nanotech. 10, 125 (2015).
[Hammel2007] Hammel and Pelekhov, The Magnetic Resonance Microscope (Handbook of Magnetism and Advanced Magnetic Materials), vol. 5 Spintronics and Magnetoelectronics, ed Kronmüller and Parkin, New York: Wiley, ISBN: 978-0-470-02217-7.
[Hansma1987] Hansma and Tersoff, J. Appl. Phys. 61, R1 (1987).

[Hartmann1962] Hartmann and Hahn, Phys. Rev. 128, 2042 (1962).
[Heinze2000] Heinze et al., Science 288, 1805 (2000).
[Herzog2014] Herzog et al., Appl. Phys. Lett. 105, 043112 (2014).
[Huang2009] Huang et al., Annu. Rev. Biochem. 78, 993–1016 (2009).
[Issac2016] Issac et al., Phys. Chem. Chem. Phys. 18, 8806 (2016).
[Jelezko2002] Jelezko et al., Appl. Phys. Lett. 81, 2160 (2002).
[Kaiser2007] Kaiser et al., Nature 446, 522 (2007).
[Kawai2005] Kawai et al., Appl. Phys. Lett. 87, 173105 (2005).
[Kawai2009] Kawai et al., Phys. Rev. B 79, 195412 (2009).
[Kawai2010] Kawai et al., Phys. Rev. B 81, 085420 (2010).
[Kelly2007] Kelly and Miller, Rev. Sci. Instrum. 78, 031101 (2007).
[Kirtley1995] Kirtley et al., Appl. Phys. Lett. 66, 1138 (1995).
[KuehnJCP2006] Kuehn et al., J. Chem. Phys. 128, 052208 (2006)
[KuehnPRL2006] Kuehn et al., Phys. Rev. Lett. 96, 156103 (2006).
[Labaziewicz2008] Labaziewicz et al., Phys. Rev. Lett. 101, 180602 (2008).
[Leskowitz1998] Leskowitz et al., Solid State Nucl. Magn. Reson. 11, 73 (1998).
[Li2008] Li et al., Nature Nanotech. 3, 88 (2008).
[Lin2006] Lin et al., Phys. Rev. Lett. 96, 137604 (2006).
[Longenecker2012] Longenecker et al., ACS Nano 6, 9637 (2012).
[Loretz2014] Loretz et al. Appl. Phys. Lett. 104, 033102 (2014).
[Loretz2015] Loretz et al., Phys. Rev. X 5, 021009 (2015).
[Lucic2005] Lucic et al., Annu. Rev. Biochem. 74, 833 (2005).
[Madsen2004] Madsen et al. Proc. Natl Acad. Sci. USA 101, 12804 (2004).
[Mamin2001] Mamin and Rugar. Appl. Phys. Lett. 79, 3358 (2001).
[Mamin2003] Mamin et al. Phys. Rev. Lett. 91, 207604 (2003).
[Mamin2007] Mamin etal., Nature Nanotech. 2, 301 (2007).
[Mamin2009] Mamin et al., Nano Lett. 9, 3020 (2009).
[Mamin2012] Mamin et al., Appl. Phys. Lett. 100, 013102 (2012).
[Mamin2013] Mamin et al., Science 339, 557 (2013).
[Mamin2014] Mamin et al., Phys. Rev. Lett. 100, 013102 (2012).
[Maze2008] Maze et al., Nature 455, 644 (2008).
[Mercier2017] Mercier de Lépinay et al., Nat. Nanotechnol. 12, 156 (2017).
[Moores2015] Moores et al., Appl. Phys. Lett. 106, 213101 (2015).
[Moores2016] Moores, PhD Thesis, ETH Zurich (2016).
[Moser2013] Moser et al., Nature Nanotech. 8, 493 (2013)
[Moser2014] Moser et al., Nature Nanotech. 9, 1007 (2014).
[Muller2006] Müller and Jerschow, Proc. Natl Acad. Sci. USA 103, 6790 (2006).
[Nestle2001] Nestle et al., Prog. Nucl. Magn. Res. Spectrosc. 38, 1 (2001)
[Nichol2008] Nichol et al., Appl. Phys. Lett. 93, 193110 (2008).
[Nichol2012] Nichol et al., Phys. Rev. B 85, 054414 (2012)
[NicholPRX2013] Nichol et al., Phys. Rev. X 3, 031016 (2013).
[NicholPhD2013] Nichol, PhD Thesis, University of Illinois (2013).
[Oosterkamp2010] Oosterkamp et al., Appl. Phys. Lett. 96, 083107 (2010).
[Peddibhotla2013] Peddibhotla et al., Nature Phys. 9, 631 (2013).
[Persson1998] Persson and Zhang, Phys. Rev. B 57, 7327 (1998).
[Pfeiffer2002] Pfeiffer et al., Phys. Rev. B 65 161403 (2002).

[Poggio2007] Poggio et al., Appl. Phys. Lett. 90, 263111 (2007).
[Poggio2010] Poggio and Degen, Nanotechnology 21, 342001 (2010).
[Poggio2013] Poggio, Nature Nanotech. 8, 482 (2013).
[Ramos2013] Ramos et al., Sci. Rep. 3, 3445 (2013).
[Rossi2017] Rossi et al., Nat. Nanotechnol. 12, 150 (2017).
[Rugar1990] Rugar and Hansma, Phys. Today 43, 23 (1990).
[Rugar1992] Rugar et al., Nature 360, 563 (1992).
[Rugar1994] Rugar et al., Science 264, 1560 (1994).
[Rugar2004] Rugar et al., Nature 430, 329 (2004).
[Rugar2015] Rugar et al., Nature Nanotech. 10, 120 (2015).
[Schaff1997] Schaff and Veeman, J. Magn. Reson. 126, 200 (1997).
[Shimizu2006] Shimizu and Itoh, Thin Solid Films 508, 160 (2006).
[Shlimak2001] Shlimak et al., J. Phys.: Condens. Matter 13, 6059 (2001).
[Sidles1991] Sidles, Appl. Phys. Lett. 58, 2854 (1991).
[SidlesPRL1992] Sidles, Phys. Rev. Lett. 68, 1124 (1992).
[SidlesRSI1992] Sidles et al., Rev. Sci. Instrum. 63, 3881 (1992).
[Sidles1993] Sidles and Rugar, Phys. Rev. Lett. 70, 3506 (1993).
[Sleator1985] Sleator et al., Phys. Rev. Lett. 55, 1742 (1985).
[Staudacher2013] Staudacher et al., Science 339, 561 (2013).
[StipePRL2001A] Stipe et al., Phys. Rev. Lett. 87, 277602 (2001).
[StipePRL2001B] Stipe et al., Phys. Rev. Lett. 86, 2874 (2001).
[StipePRL2001C] Stipe et al., Phys. Rev. Lett. 87, 096801 (2001).
[Subramaniam2005] Subramaniam, Curr. Opin. Microbiol. 8, 316 (2005).
[Suter2004] Suter, Prog. Nucl. Magn. Res. Spectrosc. 45, 239 (2004).
[Tao2014] Tao et al., Nature Comm. 5, 3638 (2014).
[TaoNL2015] Tao and Degen, Nano Lett. 15, 7893 (2015).
[TaoNano2015] Tao et al., Nanotechnology 26, 465501 (2015).
[TaoNC2016] Tao et al, Nat. Comms. 7, 12714 (2016).
[Taylor1974] Taylor and Glaeser, Science 186, 1036 (1974).
[Thurber2003] Thurber et al., J. Magn. Reson. 162, 336 (2003).
[Tsang2006] Tsang et al., IEEE Trans. Magn. 42, 145 (2006).
[Tsuji2004] Tsuji et al., J. Magn. Reson. 167, 211–20 (2004).
[Vasyukov2013] Vasyukov et al., Nature Nanotech. 8, 639 (2013).
[Verhagen2002] Verhagen et al., J. Am. Chem. Soc. 124, 1588 (2002).
[Volokitin2003] Volokitin and Persson, Phys. Rev. Lett. 91, 106101 (2003).
[Volokitin2005] Volokitin and Persson, Phys. Rev. Lett. 94, 086104 (2005).
[Wago1996] Wago et al., J. Vac. Sci. Technol. B 14, 1197 (1996).
[Wago1997] Wago et al., Rev. Sci. Instrum. 68, 1823 (1997).
[WagoAPL1998] Wago et al., Appl. Phys. Lett. 72, 2757 (1998).
[WagoPRB1998] Wago et al., Phys. Rev. B 57, 1108 (1998).
[Wigen2007] Wigen et al., Ferromagnetic resonance force microscopy Topics in Applied Physics—Spin Dynamics in Confined Magnetic Structures III vol 101, Berlin: Springer, pp 105–36 (2006).
[Wrachtrup1993] Wrachtrup et al., Nature 363, 244 (1993).
[Xiao2004] Xiao et al., Nature 430, 435 (2004).
[XueAPL2011] Xue et al., Appl. Phys. Lett. 98, 163103 (2011).
[XuePRB2011] Xue et al., Phys. Rev. Lett. 98, 163103 (2011).

[Zhang1996] Zhang et al., Appl. Phys. Lett. 68, 2005 (1996).
[Zuger1993] Züger and Rugar  Appl. Phys. Lett. 63, 2496 (1993).
[Zuger1994] Züger and Rugar J. Appl. Phys. 75, 6211 (1994).
[Zuger1996] Züger et al., J. Appl. Phys. 79, 1881 (1996).
[ZuritaSanchez2004] Zurita-Sanchez et al., Phys. Rev. A 69, 022902 (2004).